\documentclass[useAMS,usegraphicx]{mn2e}

\newcommand{\gal}{\sc morgana}
\newcommand{\gs}{\sc grasil}
\newcommand{\be}{\begin{equation}}
\newcommand{\ee}{\end{equation}}
\newcommand{\bea}{\begin{eqnarray}}
\newcommand{\eea}{\end{eqnarray}}

\def\lesssim{\,\lower2truept\hbox{${<\atop\hbox{\raise4truept\hbox{$\sim$}}}$}\,}
\def\gtrsim{\,\lower2truept\hbox{${>\atop\hbox{\raise4truept\hbox{$\sim$}}}$}\,}

\title[Dust modeling in SAMs] {Evaluating and Improving Semi-analytic modelling of Dust in Galaxies based on Radiative Transfer Calculations}

\author[Fontanot et al.]{
\parbox[t]{\textwidth}{
Fabio Fontanot$^1$,
Rachel S. Somerville$^1$,
Laura Silva$^2$,
Pierluigi Monaco$^{3,2}$,\\
Ramin Skibba$^1$
}
\vspace*{6pt}\\
$^1$MPIA Max-Planck-Institute f{\"u}r Astronomie, K{\"o}nigstuhl 17, 69117 Heidelberg, Germany\\
$^2$INAF-Osservatorio Astronomico, Via Tiepolo 11, I-34131 Trieste, Italy \\
$^3$Dipartimento di Astronomia, Universit\`a di Trieste, via Tiepolo 11, 34131 Trieste, Italy \\
email: fontanot@mpia.de}

\begin{document}
\date{Accepted ... Received ...}

\maketitle

\begin{abstract}
  The treatment of dust attenuation is crucial in order to compare the
  predictions of galaxy formation models with multiwavelength
  observations. Most past studies have either used simple analytic
  prescriptions or else full radiative transfer (RT) calculations.
  Here, we couple star formation histories and morphologies predicted
  by the semi-analytic galaxy formation model {\gal} with RT
  calculations from the spectrophotometric and dust code {\gs} to
  create a library of galaxy SEDs from the UV/optical through the far
  Infrared, and compare the predictions of the RT calculations with
  analytic prescriptions. We consider a low and high redshift sample,
  as well as an additional library constructed with empirical,
  non-cosmological star formation histories and simple (pure bulge or
  disc) morphologies. Based on these libraries, we derive fitting
  formulae for the effective dust optical depth as a function of
  galaxy physical properties such as metallicity, gas mass, and
  radius. We show that such fitting formulae can predict the $V$-band
  optical depth with a scatter smaller than 0.4 dex for both the low
  and high redshift samples, but that there is a large
  galaxy-to-galaxy scatter in the shapes of attenuation curves,
  probably due to geometrical variations, which our simple recipe does
  not capture well. However, our new recipe provides a better
  approximation to the {\gs} results at optical wavelength than
  standard analytic prescriptions from the literature, particularly at
  high redshift.
\end{abstract}


\section{Introduction}
\label{section:introduction}

Infrared (IR) observations of galaxies have demonstrated conclusively
that dust is a fundamental and ubiquitous component of the
interstellar medium (ISM). Dust grains modify the chemical and
physical conditions of the ISM by locking up a large fraction of heavy
elements ejected by stars, by shielding molecular clouds (MCs) from
dissociating radiation, thereby allowing them to cool and condense to
the densities necessary to form stars, and by favoring the formation
of $H_2$ molecules themselves on their surfaces (see e.g. the reviews
by Dorschner \& Henning 1995; Draine 2003). Therefore dust grains are
a fundamental ingredient in the very process of star formation. Also,
dust plays a major role in shaping the spectral energy distribution
(SED) of galaxies: dust grains absorb and efficiently scatter short
wavelength ($\lambda \lesssim 1 \mu$m) radiation. The extinction
efficiency drops steeply for longer wavelength and dust can not
survive at temperatures $\gtrsim 1000$K: the absorbed energy is
thermally emitted in the IR.  Therefore dust modifies the intrinsic
(pure stellar and/or non-thermal) SED of galaxies. This effect is
particularly strong in star-forming regions and starburst galaxies,
where the youngest stars are deeply embedded within dense and
optically thick dusty cocoons.

The expected link between the intensity of the star formation activity
and the amount of dust reprocessing in galaxies has been clearly
observed both in the local and in the high redshift universe,
revealing that much, and in many cases most, of the star formation
activity is obscured in the Ultra-violet (UV) and optical, and can
only be detected in the IR. For example, estimates based on
observations with the IRAS satellite suggested that globally $\sim
30$\% of the bolometric luminosity of nearby galaxies, mostly normal
spirals, is reprocessed by dust in the IR (Soifer \& Neugebauer 1991;
Popescu \& Tuffs 2002). But IRAS also revealed the existence of a
population of heavily obscured luminous and ultra-luminous IR galaxies
(with L$_{IR} \sim 10^{11} - 10^{12}$ L$_\odot$, and with L$_{IR}
\gtrsim 10^{12}$ L$_\odot$, respectively, e.g. Sanders \& Mirabel
1996), and provided a first hint of the strong evolution of these IR
bright galaxies. This has been confirmed and quantified to high
redshifts with surveys with ISO ($z \sim 0.5-1$, Elbaz et al.  1999,
2002; Gruppioni et al. 2002; Dole et al. 2001), SCUBA ($z \sim 2$;
Smail et al. 1997, 2002; Hughes et al. 1998; Chapman et al. 2005) and
{\it Spitzer} ($z \gtrsim 2$; Le Floch et al. 2005; Babbedge et al.
2006).  The optical to sub-millimetric cosmic background radiation
provides a measure of the integrated star formation activity taking
place at high-z. The energy density of the far-infrared (FIR)
background measured by COBE (Puget et al. 1996; Hauser et al. 1998;
Hauser \& Dwek 2001) is comparable to that measured in the optical and
near-IR (NIR).  All these observations indicate that the amount of
energy emitted by dust over the history of the Universe is at least
comparable to the energy emitted by stars that is able to escape
galaxies and reach us, and that star formation activity and the
consequent dust reprocessing is much more important at high redshift
than locally.

Therefore the presence and properties of dust in the ISM of galaxies,
and its effect on the SEDs must be considered with care in order to
interpret observations to infer basic quantities such as star
formation rates (SFR) and masses, as well as to compute reliable
predictions (SEDs, luminosity functions, galaxy counts etc) from
galaxy formation models.  The various spectrophotometric codes ({\sc
  pegase}, Fioc \& Rocca Volmerange, 1997; {\gs}, Silva et al., 1998;
{\sc starburst99}, Leitherer et al., 1999; Bruzual \& Charlot, 2003)
show reasonable agreement in predictions of the intrinsic, pure
stellar SED of a galaxy, given its star formation rate and metal
enrichment history. However, the SED emerging from galaxies is the
result of a complex interaction between the intrinsic properties of
dust and the relative geometry of the heating sources and the dust.
The chemical composition, size distribution, and shape of the dust
grains determine the absorption and scattering efficiencies of the
dust mixture as a function of wavelength. These properties are
somewhat constrained only for our galaxy and the Magellanic Clouds
through measurements of the extinction curve and the local diffuse
dust emission. But it is also well known that these extinction curves
have different shapes, a fact generally ascribed to different dust
properties, and also within each galaxy the curves are spatially
variable and dependent on the particular dusty environments sampled by
the line of sight (e.g.  Mathis et al. 1983; Rowan-Robinson 1986;
Cardelli et al. 1989; Mathis 1990; Fitzpatrick 1999, 2007). A further
issue is the degeneracy among different dust models that reproduce the
average extinction curve and cirrus emission in the MW (e.g. Draine \&
Anderson 1985; Desert et al 1990; Dwek et al. 1997; Li \& Draine 2001;
Zubko et al. 2004).

The relative geometry of stars and dust plays a major role in shaping
the SED. This holds both for the attenuated starlight (e.g.  Bruzual
et al. 1988; Witt et al. 1992; Efstathiou \& Rowan-Robinson 1995;
Gordon et al. 1997; Ferrara et al. 1999; Calzetti 2001) and for the
dust emission spectrum. In fact, a fundamental property to be taken
into account is that the relative geometry of stars and dust is age-
and therefore wavelength--dependent, because the youngest stars, which
dominate the UV luminosity, are also the most extinguished by the
optically thick parent MCs (Silva et al. 1998; Granato et al.  2000;
Tuffs et al. 2004). In addition, detailed reproductions of the
attenuation properties of disc galaxies seem to require an
age-dependent extinction also in the diffuse medium for intermediate
age with respect to older stars (Popescu et al. 2000; Panuzzo et al.
2007). All these factors then determine the temperature distribution
of the dust grains, with each dusty environment and each dust grain
having their own particular response to the radiation field, and
therefore yielding the consequent shape of the emerging IR SED.

In order to take full advantage of the wealth of multi-wavelength
observations now available to us (UV/optical and IR), galaxy formation
models must grapple not only with the modeling of stellar
populations, but also with absorption and re-emission by dust. This is
equally true for both semi-analytic and numerical hydrodynamic
simulations. In semi-analytic models (SAMs; e.g. White \& Frenk 1991),
the evolution of the dark matter (DM) component is calculated directly
using N-body methods or Monte Carlo techniques, while the evolution of
the baryonic component is treated by simple recipes for the radiative
cooling of gas, star formation, chemical evolution, feedback by
supernovae and AGN, etc (see Baugh 2006 for a review). The SAM
provides detailed information about the star formation and enrichment
history of each galaxy, but typically only very limited information
about the structural properties of the stars, gas, and dust (e.g., in
general an effective radius can be computed for the disc and spheroid
components).

Due to the complexities inherent to the treatment of the radiative
effects of dust and the many unknowns connected to the dust
properties, most SAMs have made use of simple empirical or
phenomenological treatments. A widely adopted approach consists in
computing the face-on dust optical depth at a reference wavelength
(typically the $V$-band, $\tau_{V}$), and then computing the
inclination dependence assuming that the stars and dust are uniformly
mixed in a ``slab'' model, and that the wavelength dependence is given
by a fixed ``template'' attenuation curve. Some modelers (e.g.
Kauffmann et al.  1999; Somerville \& Primack 1999; Nagashima et al.
2001; Mathis et al.  2002; De Lucia, Kauffmann, \& White 2004; Kang et
al. 2005) compute $\tau_{0, V}$ using empirical relations between
galaxy luminosity and dust optical depth (Wang \& Heckman 1996).
Others assume that $\tau_{0, V}$ is proportional to the column density
of dust in the disc, assuming that the dust mass is proportional to
the metallicity (Guiderdoni \& Rocca-Volmerange 1987; Lacey et al.
1993; Guiderdoni et al. 1998; Devriendt \& Guiderdoni 2000; Hatton et
al.  2003; Blaizot et al. 2004; Kitzbichler \& White 2007). De Lucia
\& Blaizot (2007) adopt the latter prescription for obscuration due to
the diffuse ``cirrus'' component, coupled with the approach suggested
by Charlot \& Fall (2000) to account for the larger obscuration of the
youngest stars by the molecular birth clouds. Cole et al. (2000) and
Bell et al. (2003) instead couple their SAM with the Ferrara et al.
(1999) library, which provides the net attenuation for smooth
distributions of stars and dust as a function of wavelength and
inclination angle based on Radiative Transfer (RT) calculations.

In some works, the IR properties of galaxies in semi-analytic models
are also computed. Using one of the above methods for computing the
optical depth and attenuation curve, and hence the total amount of
energy absorbed by the dust, and assuming that all of this light is
re-radiated in the IR, one can use observationally-calibrated
templates describing the wavelength dependence of the dust emission
(Guiderdoni et al. 1998; Devriendt, Guiderdoni, \& Sadat 1999;
Devriendt \& Guiderdoni 2000; Hatton et al. 2003; Blaizot et al.
2004), or modified Planck functions (e.g.  Kaviani et al. 2003) to
compute IR luminosities.  In a few cases, SAMs have instead been
coupled with a self-consistent RT calculation for the full (UV to
sub-mm) SED, e.g.  Granato et al.  (2000); Baugh et al. (2005), and
Lacey et al. (2007, {\sc galform}); Fontanot et al. (2007, {\sc
  morgana}); Granato et al.  (2004) and Silva et al. (2005).

Clearly there is a large range in both complexity and computation time
represented in the treatments of dust in galaxy formation models in
the literature. On the one hand, it is natural to be dubious that a
simple empirical approach can adequately represent the intricacies of
this very messy process. On the other hand, one of the advantages of
SAMs is their flexibility and computational efficiency, and this
advantage is substantially undermined if one chooses to do full RT
calculations for each galaxy. Moreover, given that SAMs are anyway
based on a series of approximations, and that so many details of the
relevant physics of galaxy formation (structural properties, star
formation, chemical evolution, supernova and AGN feedback) are poorly
understood and crudely modeled, feeding the results of SAMs into full
RT may not be well motivated.

In this paper, we make use of libraries of RT calculations based on
SAM outputs as well as sets of non-cosmological, empirical star
formation/enrichment histories to address the following questions: (1)
How different are the results from the simple analytic dust models
and the full RT calculations? (2) How do the predictions from the RT
calculations depend on the physical properties of the model galaxies?
(3) Can we derive an improved analytic recipe, based on the RT
calculations, that can be efficiently implemented in SAMs?  The SAM
that we use for this study is the {\sc morgana} model (Monaco et
al. 2007, Fontanot et al.  2006, 2007), and the RT libraries are
constructed using the {\sc grasil} code (Silva et al.  1998; Silva
1999).

{\gal} produces good overall agreement with observations of galaxy
properties; any discrepancy is of little concern in this context,
since in principle we could proceed by simply comparing SEDs resulting
from different dust prescriptions applied to any possible realizations
of the SAM. However, we wish to disentangle the dependence of the dust
extinction predictions on galaxy physical properties from the
correlations between these properties, and from distributions of these
properties characteristic of a given cosmic epoch, as predicted by
{\gal} or present in the real Universe.  Therefore we construct an
``empirical'' library, based on a classical ``open box'' chemical
evolution code with simple parameterized star formation histories,
chosen to span the broadest range of possibilities. In addition, we
consider both high-redshift and low-redshift catalogs extracted from
the {\gal} outputs. In fact we may expect that locally calibrated
simple dust prescriptions will give acceptable results when coupled to
a model characterized by a globally good agreement with local galaxy
populations. Instead this may not be the case at high-z, where
galaxies will typically have properties different from local ones. We
focus on predictions of commonly used quantities in galaxy formation
studies, namely rest-frame magnitudes in the near-UV through
Near-Infrared (SDSS $ugri$ and 2MASS $K$-band). IR luminosities from
the mid-IR (3.5 $\mu$m) to the sub-mm will be the focus of a companion
paper (Fontanot et al., in preparation).

The paper is organized as follows: in Section~\ref{section:model} we
describe the main features of the {\gal} and {\gs} models and the
construction of the RT-based libraries that we will use for our
comparison. In Section~\ref{section:ricette} we define some useful
notation and summarize the analytic prescriptions for dust attenuation
that we will evaluate. In Section~\ref{sec:physical}, we derive
fitting formulae between physical galaxy properties and the dust
effective optical depth. In Section~\ref{section:compattenu}, we
present our analysis of the results of the RT calculations in terms of
physical properties, and a comparison of the RT predictions at
UV/optical/NIR wavelengths with several analytic prescriptions.  We
give our conclusions in Section~\ref{final}. Throughout this paper we
use magnitudes in the AB system (unless otherwise stated).


\section{Galaxy Models and RT Libraries}
\label{section:model}

\begin{table*}
\begin{center}
  \begin{tabular}{cccc}
    \hline
    Sample Name & Star Formation History & Geometry & Redshift interval \\
    \hline
    Morgana Library - Disc-dominated - Low-z & {\gal} & Disc+Bulge & $0.0<z<0.2$ \\
    Morgana Library - Bulge-dominated - Low-z & {\gal} & Disc+Bulge & $0.0<z<0.2$ \\
    Morgana Library - Disc-dominated - High-z & {\gal} & Disc+Bulge & $2.0<z<3.0$ \\
    Morgana Library - Bulge-dominated - High-z & {\gal} & Disc+Bulge & $2.0<z<3.0$ \\
    \hline
    Control Library - Disc-dominated - Low-z & {\gal} & Pure Disc & $0.0<z<0.2$ \\
    Control Library - Bulge-dominated - Low-z & {\gal} & Pure Bulge & $0.0<z<0.2$ \\
    Control Library - Disc-dominated - High-z & {\gal} & Pure Disc & $2.0<z<3.0$ \\
    Control Library - Bulge-dominated - High-z & {\gal} & Pure Bulge & $2.0<z<3.0$ \\
    \hline
    Empirical Library - Disc-dominated & {\sc che\_evo} & Pure Disc & --- \\
    Empirical Library - Bulge-dominated & {\sc che\_evo} & Pure Bulge & --- \\
    \hline
  \end{tabular}
  \caption{Summary of RT libraries used in our comparison.
}
  \label{tab:libraries}
\end{center}
\end{table*}

The required inputs to the RT-based {\gs} model are the unextinguished
stellar SED and the geometry and relative distribution of stars and
dust. We assume that galaxies can be represented by a simple composite
geometry of spheroid plus disc, and that both components are
axisymmetric. The spheroid is represented by a King profile, and the
disc by a radial and vertical exponential profile. The stellar SED is
computed by convolving the distribution of stellar ages and
metallicities arising from a given star formation and enrichment
history with simple stellar population (SSP) models (Bressan, Granato,
\& Silva 1998; Bressan et al. 2002). The star formation histories, and
hence stellar SEDs, are tracked separately for the bulge and disc
components. We describe the details of how these quantities are
obtained for each of our Libraries in the following sections.

We make use of two different approaches to obtain ensembles of these
input parameters. The first approach is based on semi-analytic
simulations with the {\gal} code (details given in
\S\ref{section:morgana} below) within the $\Lambda$CDM cosmological
context. We extract catalogs at both low and high redshift from
{\gal}. We refer to the libraries that are obtained from the {\gal}
outputs as the {\it Morgana Library} (ML). In addition, we create a
library in which the star formation/enrichment histories are predicted
by {\gal} as before, but instead of having a composite disc+spheroid
morphology, disc-dominated galaxies are represented as a pure disc and
bulge-dominated galaxies as a pure bulge. In this way we can attempt
to disentangle the effects of geometry and star formation history. We
refer to this as the {\it Control Library} (CL).

For the second approach, we use a classical chemical evolution code
({\sc che\_evo}), which is not embedded in a cosmological context
(details are given in \S\ref{section:cheevo}). The goal is to remove
the ``prior'' that is contained in the cosmological models, in the
form of the predicted correlations between and distributions of
physical parameters. We refer to this as EL ({\it Empirical Library}).

All libraries assume a Salpeter (1955) IMF with mass range from $0.1$
to $100$ M$_\odot$.  Table~\ref{tab:libraries} summarizes the
different libraries considered in this paper and described in this
Section.

\subsection{Semi-analytic galaxy formation model: {\sc morgana}}
\label{section:morgana}

\begin{figure*}
  \centerline{
    \includegraphics[width=9cm]{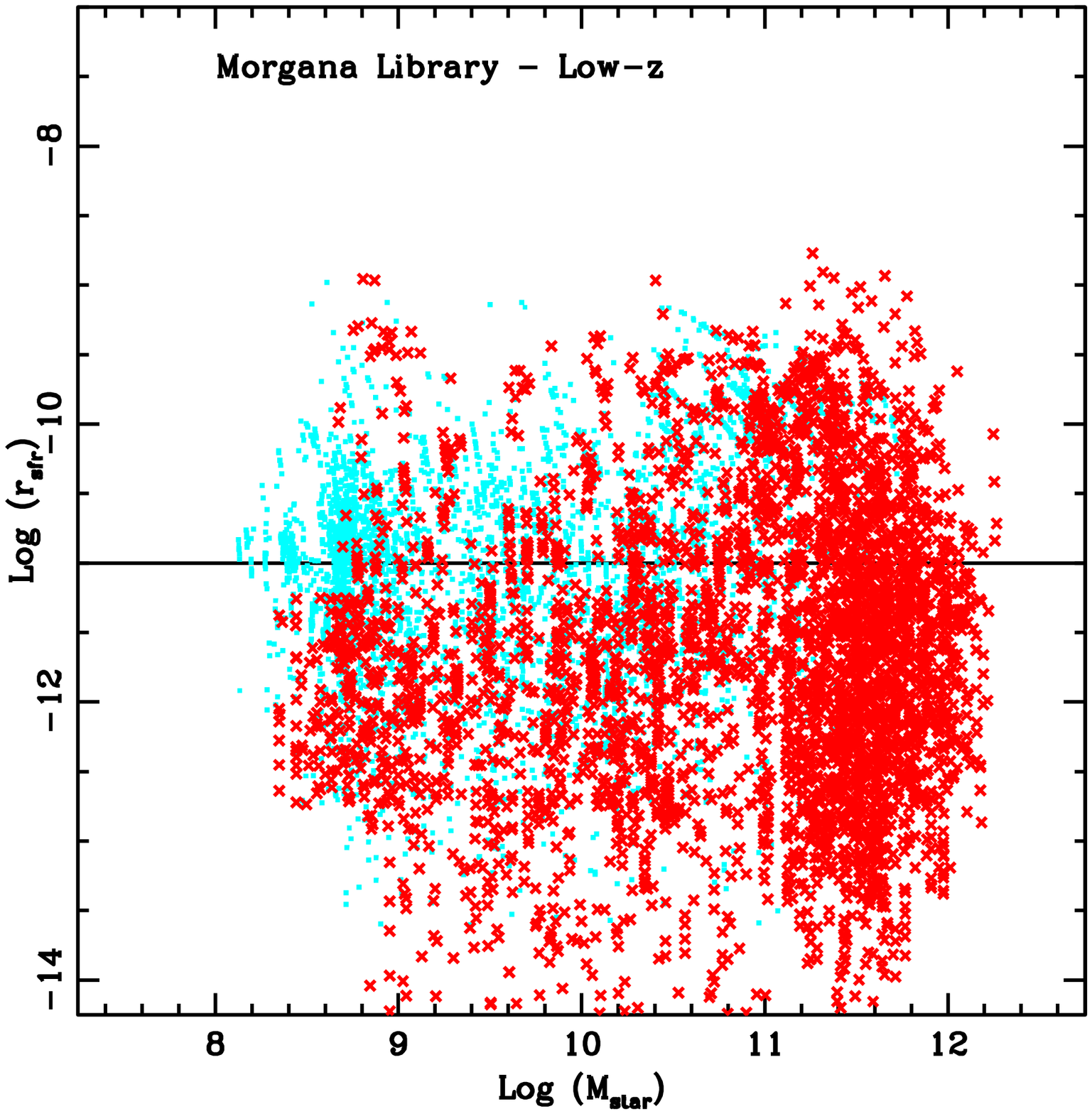}
    \includegraphics[width=9cm]{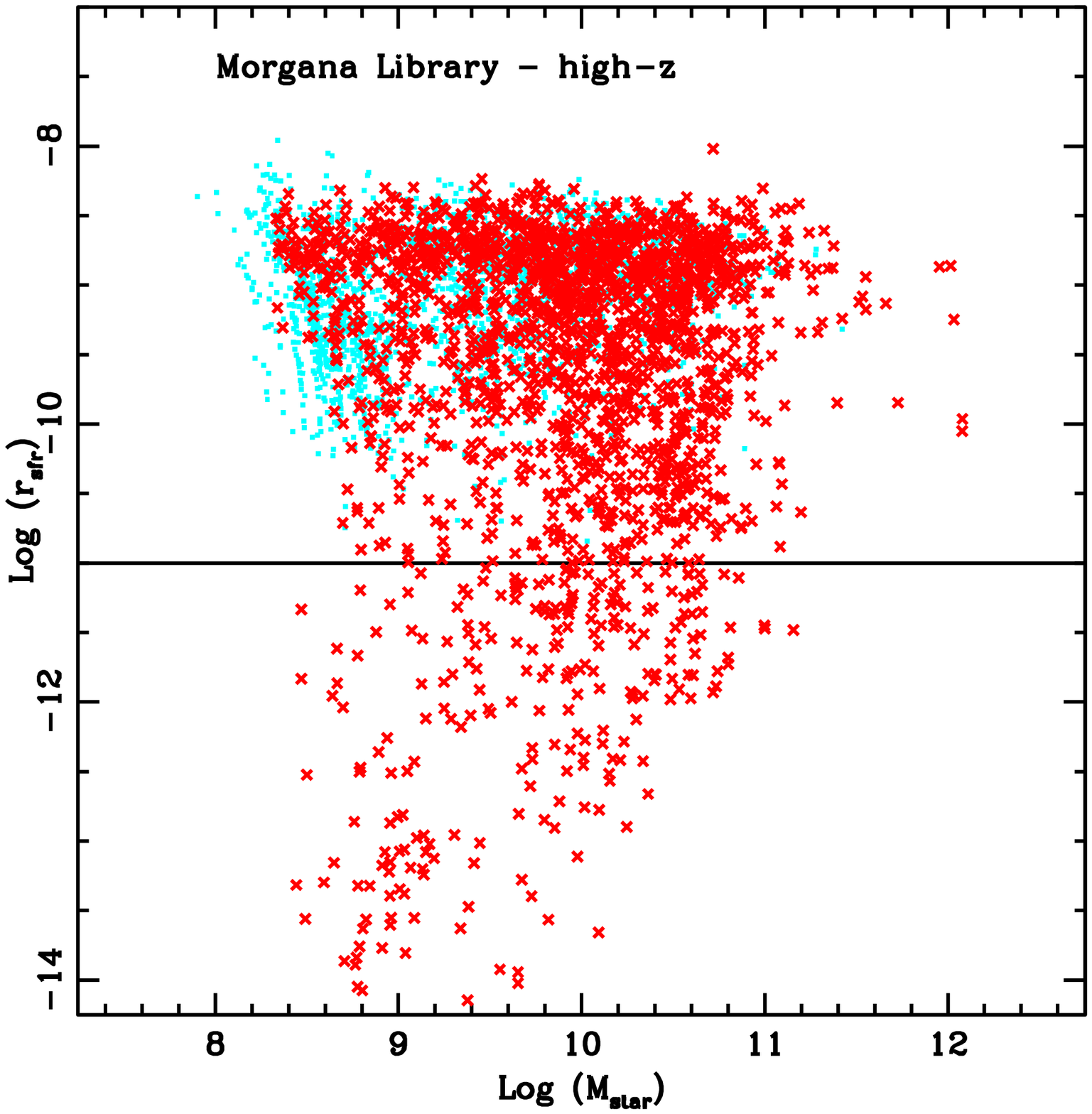} }
  \caption{$\log(M_\star/M_\odot)$ versus $\log(r_{sfr}/yr^{-1})$ for
    the {\gal} library (ML). Red crosses and cyan circles refer
    respectively to bulge-dominated and disc-dominated objects. Left
    and right panel refer to the low-z and high-z sample
    respectively.}
  \label{svm2}
\end{figure*}

{\gal} (MOdel for the Rise of GAlaxies aNd Active nuclei) is a
semi-analytic model for the formation and evolution of galaxies and
AGNs (Monaco, Fontanot \& Taffoni 2007; Fontanot et al. 2006, 2007).
The main characteristics and ingredients it implements are the
following: (i) a sophisticated treatment of the mass and energy flows
between galactic phases (cold and hot gas, stars) and components
(bulge, disc and halo) coupled with the multi-phase treatment of the
interstellar and intra-cluster media; (ii) an improved modeling of
cooling and infall (Viola et al.  2007); (iii) a multi-phase
description of star formation and feedback processes (Monaco et
al. 2004); (iv) a self-consistent description of AGN activity and
feedback (Fontanot et al. 2006); (v) the building of the diffuse
stellar component in galaxy cluster from the stars scattered in galaxy
mergers (Monaco et al.  2006).

The {\gal} realization from which we extract the library of
star-formation histories considered here, the ML, is the same
presented in Fontanot et al. (2007), and we refer to the that paper
for more details. It shows good agreement with the local stellar mass
function, cosmic star formation history, the evolution of the stellar
mass density, the slope and normalization of the Tully-Fisher relation
for spiral discs, the redshift distributions and luminosity function
evolution of $K$-band selected samples, and the $850 \mu$m number
counts. Despite the fact that this model is able to reproduce the
overall assembly of massive galaxies, it does not reproduce the
``downsizing'' trend of galaxies: in particular it overpredicts the
number of bright galaxies at $z<1$ (Monaco et al. 2006) and the number
of faint galaxies at $z \ge 1$ (Fontana et al. 2006). Moreover the
model still underpredicts the counts of the brightest $850 \mu$m
galaxies at high-z. In this model we neglect other ingredients, such
as quasar-triggered galactic winds (see Monaco \& Fontanot 2005). We
checked that including these ingredients does not change the
conclusions presented in this paper.

{\gal} gives predictions for the star formation histories of disc and
bulge components separately, and for galaxy sizes. Disc sizes are
computed using the Mo, Mao \& White (1998) approach. The spin
parameter of the DM halo is randomly chosen from its known
distribution, and it is assumed that the angular momentum is
conserved. The presence of a central bulge is taken into account in
computing the disc size.  Bulge sizes are computed (as, e.g., in Cole
et al. 2000) by assuming that kinetic energy is conserved when the
merger of two collisionless discs takes place. We classify the objects
in the ML into two categories according to the stellar {\it
  bulge-to-total} mass ratio $B/T =
\frac{M_{bulge}}{M_{bulge}+M_{disc}}$.  Following Magdwick et al.
(2003) we assume bulge-dominated galaxies to correspond to $B/T >
0.6$, and disc-dominated to $B/T \le 0.6$.

We compare the physical properties of our model galaxies with the SDSS
sample, and in particular with the specific star formation $r_{sfr} =
SFR/M_{\star}$ vs. stellar mass $M_{\star}$ plot presented in
Brinchmann et al.  (2004). They find that massive galaxies show low
$r_{sfr}$ values, whereas less massive galaxies populate a horizontal
strip in the diagram, corresponding to higher $r_{sfr}$ values.
Moreover, Madgwick et al. (2003) find a relation between the
morphological type and $r_{sfr}$: objects with $\log(r_{rsf}) < -11$
($\log(r_{rsf}) > -11$) tend to be bulge- (disc-) dominated galaxies.
We show the corresponding plot for the objects in the ML in
Fig.~\ref{svm2}; we mark the position of bulge- (disc-) dominated
galaxies as red crosses (cyan circles). It is evident from this plot
that the massive objects in {\gal} favor a bulge-dominated morphology,
whereas the low-mass tail is populated by a large fraction of
disc-dominated objects, in qualitative agreement with the SDSS
observations. However, many massive galaxies show residual star
formation activity, which is not seen in the SDSS data: this is likely
due to the incomplete quenching of cooling flows at late times in
{\gal} (see Monaco et al. 2007 for a complete discussion of this
point). Moreover the star formation activity of many disc-dominated
galaxies is too low with respect to observations, in part likely due
to the strangulation processes acting on cluster satellite galaxies.
However, these residual discrepancies between observations and model
are not a major issue for this paper.

Many relevant properties of galaxies, such as the star formation rates
and sizes, evolve with redshift. We consider two samples drawn from
{\gal}, a ``low-z'' sample at $z<0.20$ and a ``high-z'' sample at
$2.0<z<3.0$. In the following we will refer to the libraries based on
these two samples as the ``low-z'' and ``high-z'' libraries. For both
libraries we define disc- and bulge-dominated sub-samples. Starting
from these samples we also define the CL accordingly.

\subsection{Simple star formation history model: {\sc che\_evo}}
\label{section:cheevo}

\begin{table}
\begin{center}
  \begin{tabular}{ccc}
    \hline
    Parameter & lower value & upper value \\
    \hline
    $t_{\rm wind}$ & 1 Gyr & 15 Gyr\\
    $t_1$ & 3 Gyr & 10 Gyr \\
    $t_2$ & $t_1$ & $t_1$ +7 Gyr \\
    $M_{\rm infall}$ & $10^7 M_\odot$ & $10^{14} M_\odot$ \\
    $\tau_{\rm infall}$ & 5 Gyr & 10 Gyr \\
    $f_\star$ & 0 & 1 \\
    $e_\star$ & 0 Gyr & 7 Gyr \\
    $\epsilon_{\rm sch-en}$ & 0 & 10 \\
    \hline
  \end{tabular}
  \caption{Parameters for the generation of random star formation
  histories with {\sc che\_evo}.}
  \label{chevo_tab}
\end{center}
\end{table}

In order to explore the correlation between dust attenuation and the
physical properties of model galaxies, we need to define a different
sort of model library. The recipes that determine galaxy evolution in
{\gal} (star formation, feedback etc.) are responsible for built-in
correlations among the physical quantities that enter in the
determination of dust attenuation (metallicity, mass of gas and/or
stars, star formation rate, radius of the galaxy, bolometric
luminosity). These recipes (and their induced correlations) differ
from SAM to SAM.  In order to make sure our results are independent of
the particular SAM we consider, we define another library, EL, using
the chemical evolution code {\sc che\_evo} (Silva 1999).

This is a ``classical open box'' chemical evolution code to compute
the star formation and metallicity evolution of galaxies, with the
following characteristics. (i) The gas from which stars form has an
infall rate given by an exponential function of time with an e-folding
time $\tau_{\rm infall}$. The normalization of the gas infall rate is
fixed by setting the total mass $M_{\rm infall}$ at a given time
$t_{\rm infall}$. (ii) The SFR can be expressed as a function of the
mass of gas, as an arbitrary function of time, or as a combination of
both: $SFR(t)=\nu_{\rm sch} \, M_{\rm gas}(t)^{k_{\rm sch}} +
f(t;M_{\rm gas})$, where the first term is a Schmidt-type law with
efficiency $\nu_{\rm sch}$ and exponent $k_{\rm sch}$.  The second
term represents a superimposed mode for which we set the initial and
final time $t_1$ and $t_2$ and the fraction of gas converted into
stars $f_\star$. We choose two possibilities for this mode: an
exponential form depending only on time with e-folding time $e_\star$,
and a ``Schmidt enhanced'' mode, with efficiency $\epsilon_{\rm
  sch-en}$ and exponent $\alpha_{\rm sch-en}$. (iii) It is possible to
set a time $t_{\rm wind}$ to represent the development of a galactic
wind after which all the gas is removed, and the SFR and infall are
stopped.  All the gas present after this age comes from stellar
evolution. (iv) The metallicity evolution $Z(t)$ is computed by
accounting for the lifetimes and yields of stars as a function of
their mass and metallicity, including also type Ia supernovae.  The
evolution of all the quantities is then computed to a final age
$t_{\rm fin}$.

In order to define a library with a uniform coverage of the possible
evolutionary histories, we randomly extract the values of most of the
{\sc che\_evo} parameters within large intervals, as reported in
Table~\ref{chevo_tab}. For each combination of those parameters we
consider $t_{\rm fin}=5$, 10, 13.6 Gyr.  We fix $t_{\rm infall}= 13$
Gyr (which sets the normalization of the infall rate), $k_{\rm
  sch}=1$, $\alpha_{\rm sch-en}=1$, since their variation does not add
further information.  In order to break possible remaining
correlations, we then renormalize separately the star-formation
history and the evolution of the cold gas mass to final values
extracted randomly from uniform distributions. This choice forces the
final metallicity evolution to be independent from the star-formation
history and the cold gas content. The final result of this procedure
is a uniform distribution of the models in the ($SFR$, $M_{\rm gas}$,
$Z$) space.  We then randomly assign simple morphologies (i.e. pure
disc or pure bulge) to the objects in the EL. We also recompute the EL
using the Madgwick et al. (2003) criterion for assigning morphologies
based on their specific star formation rates: we assume that objects
with $\log(r_{sfr}) < -11$ correspond to pure spheroidal galaxies, and
objects with $\log(r_{sfr}) \ge -11$ are considered to be pure discs.
We have checked that our conclusions do not change between the two
libraries, so we will present results only for the first realization.

\subsection{Dust and Radiative Transfer model: {\sc grasil}}
\label{section:grasil}

For each object in the {\gal} and {\sc che\_evo} libraries, we compute
the corresponding UV to radio SED using the spectrophotometric code
{\gs} (Silva et al. 1998). Subsequent updates and improvements are
described in full detail in Silva (1999); Granato et al. (2000);
Bressan, Silva \& Granato (2002); Panuzzo et al. (2003); Vega et al.
(2005). We refer to these papers for more details.

{\gs} solves the equation of RT, taking into account a
state-of-the-art treatment of dust effects, and includes the following
main features: (i) stars and dust are distributed in a bulge (King
profile) + disc (radial and vertical exponential profiles)
axisymmetric geometry; (ii) the clumping of both (young) stars and
dust through a two-phase interstellar medium with dense giant
star-forming molecular clouds embedded in a diffuse (``cirrus'') phase
are considered; (iii) the stars are assumed to be born within the
optically thick MCs and to gradually escape from them on a time-scale
$t_{\rm esc}$, this gives rise to the age- (wavelength-) dependent
extinction with the youngest and most luminous stars suffering larger
extinction than older ones; (iv) the dust composition consists of
graphite and silicate grains with a distribution of grain sizes, and
Polycyclic Aromatic Hydrocarbons (PAH) molecules; (v) at each point
within the galaxy and for each grain type the appropriate temperature
$T$ is computed (either the equilibrium $T$ for big grains or a
probability distribution for small grains and PAHs); (vi) the
radiative transfer of starlight through dust is computed along the
required line of sight yielding the emerging SED; (vii) the simple
stellar population (SSP) library (Bressan, Granato, \& Silva 1998;
Bressan et al. 2002) includes the effect of the dusty envelopes around
AGB stars, and the radio emission from synchrotron radiation and from
ionized gas in HII regions.

For the {\gal}+{\gs} library (ML or CL), the scale radii for stars and
dust in the disc ($r_d^\star$, $r_d^d$) and in the bulge ($r_b^\star$,
$r_b^d$) are provided by {\gal}. The disc scale heights for stars and
dust ($h_d^\star$ and $h_d^d$) are set to $0.1$ times the
corresponding scale radii.

For the {\sc che\_evo}+{\gs} library (EL), we assume that the scale
radius depends on stellar mass as:
\be
r = k_r \alpha_r \left ( \frac{M_\star}{10^{11} M_\odot} \right )^{1/3} {\rm kpc} \\
\ee

\noindent
where $\alpha_r = 0.2$ kpc for $r_b^\star$, $\alpha_r = 3.0$ kpc for
$r_d^\star$ and $k_r$ is a random number between 0.5 and 2. We assume
the same scale radii for the stellar and dust components and set the
disc scale-heights to $0.1$ times the corresponding scale radii.

We fixed the other parameters needed by {\gs} and not provided by the
star formation models. (i) The escape time-scale of young stars for
the parent MCs is set to $t_{\rm esc} = 10^7$ yr, an intermediate
value between that found by Silva et al. (1998) to describe well the
SED of spirals ($\sim$ a few Myr) and starbursts ($\sim$ a few 10
Myr), and of order the estimated destruction time scale of MCs by
massive stars. In Fontanot et al. (2007) this value was found to
produce good agreement with the K-band luminosity functions and the
$850 \, \mu$m counts.  (ii) The gas mass predicted by {\gal} or {\sc
  che\_evo} is subdivided between the dense and diffuse phases,
assuming the fraction of gas in the star forming molecular clouds
$f_{\rm MC}$ is $0.5$. The results are not very sensitive to this
choice. (iii) The mass of dust is obtained by the gas mass and the
dust to gas mass ratio $\delta_{\rm dust}$ which is set to evolve
linearly with the metallicity given by the galaxy model, $\delta_{\rm
  dust}=0.45 \; Z$.  (iv) The optical depth of MCs depends on the
ratio $\tau_{MC} \propto \delta_{\rm dust} \, M_{\rm MC}/r_{\rm
  MC}^2$; we set the mass and radius of MCs to typical values for the
MW, $M_{\rm MC} = 10^6 M_\odot$ and $r_{\rm MC} = 16$ pc. (iv) The
dust grain size distribution and composition is chosen to match the
mean MW extinction curve.

{\gs} computes the resulting SEDs along different lines of sight, but
in the following we use only angle-averaged SEDs, unless otherwise
stated. We convolve the SEDs with the GALEX FUV, SDSS $u$, $g$, $r$,
$i$ and 2-MASS $K_s$ filters.


\section{Analytic and empirical prescriptions for attenuation}
\label{section:ricette}

\subsection{Definitions and notation}
\label{section:def}

For clarity, we first define some useful notation and the meaning of
several quantities that we will use throughout the remainder of the
paper.

The amount and wavelength dependence of the dust attenuation on the
intrinsic (pure stellar) SED of a system is commonly defined through
the ratio between the emerging (after interaction with dust) and
intrinsic luminosity, typically expressed in magnitudes:

\be A_\lambda \equiv m_\lambda - m_\lambda^0 = -2.5 \, \log
L_\lambda/L_\lambda^0 \ee

It is important to clearly define and to explicitly distinguish
between the {\it extinction} and the {\it attenuation} curve
(e.g. Granato et al. 2000; Calzetti 2001). The term {\it extinction}
is commonly used to describe the wavelength dependence of the optical
properties (absorption plus scattering) of the dust mixture. Indeed,
in the simple geometrical configuration of a slab of dust between the
observer and a point source, as is the case when directly measuring
the extinction from observations of background stars, since the dust
emission along the line of sight (both true dust emission and
scattering) is negligible within a point source image, the solution of
the equation of RT provides a direct link between the ratio of the
observed to intrinsic energy, and the dust optical depth:
$L_\lambda/L_\lambda^0=\exp(-\tau_\lambda)$.  Therefore in this case
$A_\lambda=1.086 \, \tau_\lambda$, where
$\tau_\lambda=\tau_{\lambda,abs}+\tau_{\lambda,sca}$ contains the
information on the dust properties along the line of sight.

This simple situation of course does not apply to studies of external
galaxies. In this case we refer to $A_\lambda$ as the {\it
  attenuation} curve (other commonly used terms with the same meaning
are obscuration, effective extinction or absorption), since in this
case the solution of the RT equation does not provide a simple nor a
priori predictable relation for $L_\lambda/L_\lambda^0$. For
convenience, it is common practice to define an ''effective'' optical
depth such that $L_\lambda/L_\lambda^0=\exp(-\tau_\lambda^{eff})$,
with $\tau_\lambda^{eff}$ summarizing the complex interaction between
stars and dust in galaxies. This is further discussed in
\S\ref{section:attcurve}.

In the following, we refer to these quantities, the attenuation and
the corresponding effective optical depth obtained with {\gs} as
$A_\lambda^{GS}$, $\tau_\lambda^{GS}$, and the color excess as
\be E^{GS}(\lambda_1-\lambda_2)=A_{\lambda_1}^{GS}-A_{\lambda_2}^{GS}.
\ee
We will compare the values of these quantities obtained from {\gs}
with those that we obtain when we apply various analytic dust
prescriptions to the pure stellar SED.

\subsection{Analytic prescriptions for attenuation}
\label{section:attenu}

Several simple and computationally efficient treatments of dust
effects in galactic SEDs have been proposed. They adopt different
approaches to the problem, making a direct comparison
difficult. However they also share some basic ingredients : (1) an
attenuation law, either empirical (e.g. Calzetti et al. 1994; 2000) or
obtained by coupling an extinction law with an idealized geometrical
configuration (2) a face-on effective optical depth at a reference
wavelength (usually $V$-band), (3) a geometrical model for the
dependence of the attenuation on viewing angle. Below we describe some
of the most commonly adopted prescriptions for these ingredients,
focusing on those that we will compare with in the next section.

\subsection{Extinction and Attenuation Laws}

For our own and a few nearby galaxies, the extinction law of the dust
has been directly measured from observations of background stars.  The
differences found between the shapes of the (average) extinction
curves of the Galaxy, the Large Magellanic Cloud and the Small
Magellanic Cloud below $\sim 2600$ \AA\ are often ascribed to the
different metalicities in these systems. The Galactic extinction curve
for the diffuse medium is commonly adopted (e.g.  Mathis et al. 1983;
Fitzpatrick 1989).

Calzetti et al. (1994, 2000) have analyzed the dust attenuation for a
sample of UV-bright starbursts.  The derived attenuation curve is
characterized by a shallower UV slope than that of the MW extinction,
and by the absence of the $2175$ \AA \, feature. A complementary
method of defining an empirical attenuation law has been presented by
Charlot \& Fall (2000, hereafter CF00). They consider an attenuation
curve that is a power-law function of wavelength, with a normalization
that depends on the age of the stellar population: \be
\tau_\lambda^{CF} = \left\{
\begin{array}{ll}
\tau_V (\lambda/5500 \AA)^{-0.7}  & \textrm{if $t \le t_{MC}$} \\
\mu \, \tau_V (\lambda/5500 \AA)^{-0.7} & \textrm{if $t > t_{MC}$} \\
\end{array}
\right.
\label{cf}
\ee
\noindent
where $\tau_V$ indicates the total V-band attenuation experienced by
young stars within the birth MCs (due to the birth clouds themselves
and the diffuse ISM), $t_{MC}$ is the timescale of destruction of the
MC or for stars to migrate out of the birth clouds and $\mu$ defines
the fraction of the total effective optical depth contributed by the
ambient ISM. CF00 used a sample of nearby UV-selected starburst
galaxies to test their model and constrain its parameters, by
considering the ratio of FIR to UV luminosities, the UV spectral
slope, and Balmer line ratios (see also Kong et al. 2004).  The
best-fit values found by CF00 are $t_{MC} \simeq 10^7 yr$, $\mu \sim
0.3$, with $\tau_V^{ISM}=\mu \, \tau_V \sim 0.5$ and $\tau_V^{MC} =
(1-\mu) \, \tau_V \sim 1.0$.

In the spirit of the CF00 model, one can of course adopt a composite
attenuation curve, e.g. in which the Milky Way extinction curve is used
to describe the diffuse ``cirrus'' dust and the CF00 power-law
attenuation law is used to describe the situation experienced by young
stars in the dense birthclouds. This is the approach taken by De Lucia
\& Blaizot (2007, DLB07 hereafter).

\subsection{Effective Optical Depth}

Perhaps the most fundamental quantity characterizing the extinction is
the {\it face-on effective optical depth} in a given reference band,
usually the V band ($\tau_V$). As we discussed in the introduction,
many works have made use of the empirical relation between the
intrinsic luminosity in the B-band and the optical depth (Wang \&
Heckman 1996). Also very common are variants of the model originally
proposed by Guiderdoni \& Rocca-Volmerange (1987, hereafter GRV87),
which links the optical depth of the dust at a given wavelength
$\tau_\lambda$ to the column density of metals in the cold gas and to
$\tau_V$:

\be \tau_\lambda = \left ( \frac{A_\lambda}{A_V} \right ) \left (
  \frac{Z_{gas}}{Z_\odot} \right )^s \, \left( \frac{ N_H }{2.1 \times
    10 ^{21} {\rm cm}^{-2}} \, \right )
\label{grv}
\ee
\noindent
The exponent $s$ of the metallicity is wavelength dependent: $s =
1.35$ for $\lambda < 2000 \AA$, and $s = 1.6$ for $\lambda > 2000
\AA$, based on interpolations of the extinction curves of the MW and
the Magellanic Clouds. In GRV87 the column density $N_H$ is assumed to
be simply proportional to the cold gas fraction:
\be \langle N_H \rangle = 6.8 \times 10^{21} \frac{M_{\rm
    gas}}{M_\star + M_{\rm gas}} {\rm cm}^{-2}.
\label{grv2}
\ee

More recent models, in which an estimate of the radius of the disc has
been made available, have used an updated form in which the column
density is proportional to the cold gas mass divided by the gas radius
squared (e.g. Guiderdoni et al. 1998; Devriendt \& Guiderdoni 2000;
Cole et al. 2000). DLB07 assume:
\be N_H = \frac{M_{gas}}{1.4 m_p \pi \, r^2} {\rm cm^{-2}}
\label{dlb}
\ee
where $r$ is the exponential scale radius of the disc, as predicted by
their semi-analytic model.

As discussed above, the extinction suffered by young stars enshrouded
in dense birth clouds may be greater than that experienced by more
evolved stars embedded in more diffuse cirrus. We will adopt the
age-dependent extinction prescription proposed by DLB07. They assume
that stars younger than $t_{MC}$ (which they take to be $10^7$ yr),
have an optical depth $\tau_V^{MC}=\tau_V^{\rm cirrus} \, (1/\mu -
1)$, with the value of $\mu$ randomly extracted from a Gaussian
distribution with $\langle \mu \rangle = 0.3$ and $\sigma_\mu =
0.2$, truncated at 0.1 below and 1.0 above.

\subsection{Geometry}

The adopted extinction or attenuation curves are then generally
combined with simple geometrical configurations having analytical
expressions, in order to describe the dependence of the attenuation on
the viewing angle $i$ and/or include scattering. The analytical
solutions for the radiative transfer in simplified geometries are
generally taken from Natta \& Panagia (1984), Osterbrock (1989), Lucy
et al. (1991), and Varosi \& Dwek (1999).

The simplest and most commonly used assumption is the {\it slab}
configuration, where stars and dust are homogeneously distributed in
an infinite plane layer:

\be A_{\lambda,i}^{\rm slab} = -2.5 \, \log \left(
  \frac{1-\exp(-\tau_\lambda/\cos(i))}{\tau_\lambda/\cos(i)} \right)
\label{slab}
\ee

The $\tau_\lambda$ in the expression is sometimes ``corrected'' for
scattering by multiplying it by $\sqrt{1-\omega_\lambda}$, where
$\omega_\lambda$ is the albedo (Guiderdoni \& Rocca-Volmerange, 1987).
Other geometries, such as the {\it oblate ellipsoid}, have also been
considered (e.g. Devriendt et al. 1999).

\section{Relating dust extinction to physical parameters}
\label{sec:physical}

\begin{figure*}
  \centerline{
    \includegraphics[width=9cm]{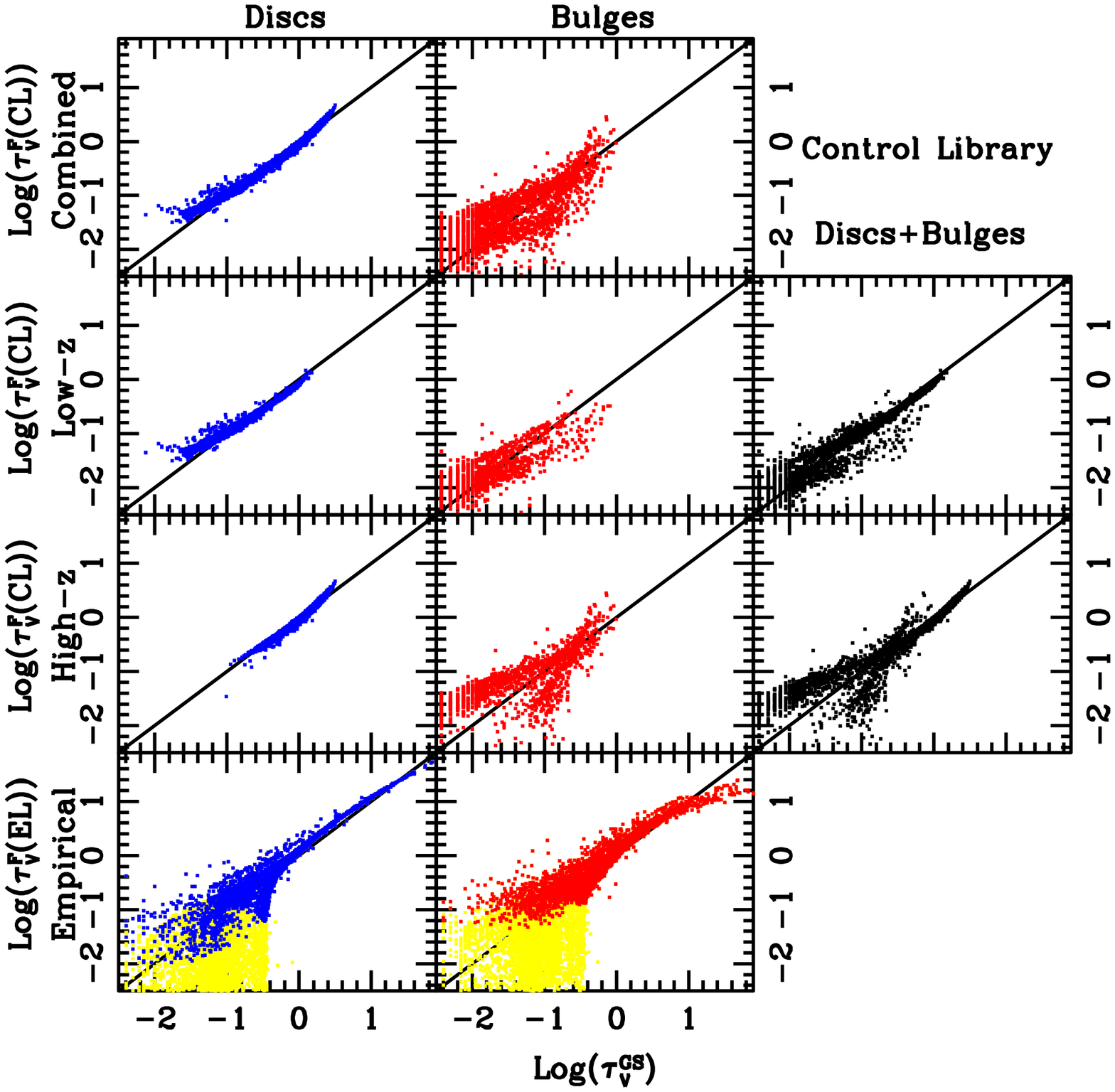}
    \includegraphics[width=9cm]{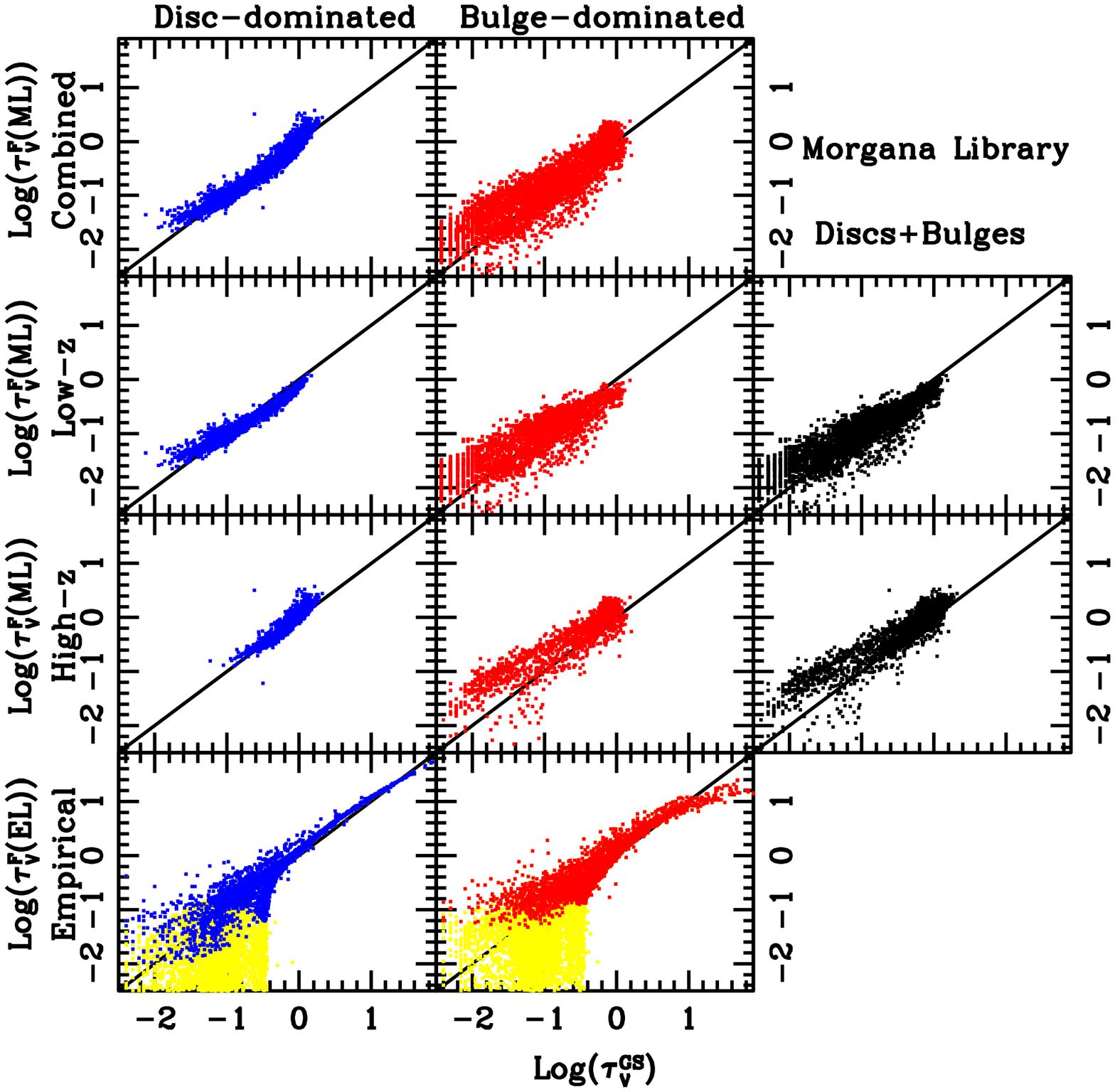}}
  \caption{Comparison between the ``true'' optical depth from the RT
    libraries $\tau_V^{GS}$ and the predictions of our best-fit
    formulae (eq.~\ref{fit2}). The bottom row shows the results from
    the Empirical Library (EL). The upper three rows show the results
    from the combined (high-z plus low-z), and separate high-z and
    low-z samples (as labeled) for the Control Library (CL; left
    plot) and the Morgana Library (ML; right plot). Vertical columns
    within each plot show the bulge-dominated, disc-dominated, and
    combined (bulge and disc) samples, as labeled. }
  \label{taufitfig}
\end{figure*}

The GRV87 model for the optical depth, and its offshoots, provided a
first basic attempt to relate the dust optical depth to physical
properties of galaxies such as metallicity and gas column density. As
we have discussed, empirically it is also known that dust optical
depth correlates with properties such as bolometric or B-band
luminosity. Based on RT calculations with the ray-tracing code {\sc
SUNRISE} (Jonsson 2006) applied within hydrodynamical simulations (Cox
et al., 2004), Jonsson et al. (2006) found a tight relationship
between effective optical depth and combinations of quantities such as
metallicity, bolometric luminosity, star formation rate, stellar mass
or baryonic mass. Motivated by these results, we analyze our libraries
to attempt to extract a generalized fit for the effective optical
depth as a power-law function of these variables, e.g.
\be \tau_V =
f(L_{\rm bol}, M_{\rm bar}, SFR, Z, M_{\rm gas}, M_\star, r_{\rm gal})
\label{eq:gen}
\ee

As we discussed in \S\ref{section:cheevo}, correlations like the one
found by Jonsson et al. (2006) may be in part the result of built-in
correlations in the models, e.g., the Schmidt law used in the
hydrodynamic simulations relates star formation rate to gas density.
Therefore, we initially explore the dependence of dust obscuration on
galaxy properties in our empirical {\sc che\_evo}+{\gs} library, EL,
where we have deliberately broken all such built-in correlations. We
then explore whether the same correlations hold in the ML.

As we describe in \S\ref{section:model}, in the EL, $Z$, $M_{\rm gas}$
and $SFR$ are uncorrelated by construction. The remaining 4 physical
quantities ($L_{\rm bol}$, $M_\star$, $M_{\rm bar}$ and $R_{\rm gal}$)
follow from the star formation history and our assumed correlation
between $M_\star$ and $R_{\rm gal}$. We then perform a fitting
procedure to try to recover the ``true'' effective optical depth from
the {\gs} library, $\tau_V^{GS}$. We computed $\tau_V^{GS}$ from the
corresponding face-on {\gs} SEDs. We assume power-law dependences for
the values of the 7 physical quantities and we define general
relations involving independent physical quantities.  We determine the
best-fit parameters in each relation using a $\chi^2$ minimization
technique and we repeat the fitting procedure considering the
disc-dominated and the bulge-dominated subsamples separately. We start
from a relation involving a single physical quantity, and then
consider an additional degree of freedom if the resulting reduced
$\chi^2$ is lowered by a factor $\sim 2$. Our results show that we
usually need to involve at least three physical quantities in order to
reproduce $\tau_V^{GS}$ well; and in particular the best fit result is
obtained by considering the following relation:

\be \tau_V^F = \tau_V^0 \bigg ( \frac{Z}{Z_\odot} \bigg )^\alpha \bigg
( \frac{M_{\rm gas}}{M_\odot} \bigg )^\beta \bigg ( \frac{R_{\rm
gal}}{1 {\rm kpc}} \bigg )^{\gamma}
\label{fit2}
\ee

\noindent
In the following we refer to the $\tau_V$ values predicted by this
fitting formula as $\tau_V^F$, and we indicate in parenthesis the
library we use to calibrate the best-fit parameters.

The optical depth $\tau_V^{GS}$ in the EL is very clearly related to
the gas mass $M_{\rm gas}$, scale radius $R_{\rm gal}$ and $Z$.  This
is not surprising, given that the dust-to-gas ratio we have used in
{\gs} is proportional to the metallicity (see \S\ref{section:grasil}).
However the relation between $\tau_V^{GS}$ and $\tau_V^F(EL)$ shows a
large spread for low values of $\tau_V^{GS}$; we checked that this
behavior is connected to low values of the gas surface density
$\Sigma_{\rm gas} = M_{\rm gas}/(4 \pi R^2_{\rm gal})$. For this
reason, when computing the best-fit parameters for the EL (see
tab.~\ref{tab:bfv}) we restrict our analysis to objects with
$\Sigma_{\rm gas}> 1 M_\odot {\rm pc}^{-2}$ for the EL discs and
$\Sigma_{gas}> 100 M_\odot {\rm pc}^{-2}$ for the EL bulges (the
different values are related to the different meaning of the scale
radius in the two geometries). The best-fit relation between
$\tau_V^{GS}$ and $\tau_V^F(EL)$ is shown in Fig.~\ref{taufitfig},
lower panels. Yellow dots refer to the whole EL, whereas blue and red
dots refer to the objects with a gas surface density higher than the
threshold.

We then consider the SEDs in the CL. Recall that the CL uses the star
formation and enrichment histories predicted by {\gal}, but with
bulge-dominated systems represented by a pure bulge and disc-dominated
systems represented by a pure disc. We see that $\tau_V^F(EL)$
provides a good approximation for $\tau_V^{GS}$ in this library, for
the disc sample. On the other hand most of the objects with bulge
geometry lie below our surface density threshold, and therefore
$\tau_V^F(EL)$ is not a good solution for this subsample: in
particular it provides a systematic underestimate of the
$\tau_V^{GS}$. We then redetermine the best-fit parameters for
eq.~\ref{fit2} (i.e. the relation involving $M_{gas}$, $R_{gal}$ and
$Z$) on the combined (low-z + high-z) CL and collect the results in
tab.~\ref{tab:bfv}. In fig.~\ref{taufitfig} (left panel) we show the
comparison between $\tau_V^{GS}$ and $\tau_V^F(CL)$: for the pure
discs the fit is accurate both for the combined and for the two
separate samples; for the pure bulges the quality of the fit is
slightly different for the ``low-z'' and ``high-z'' samples.

\begin{table*}
\begin{center}
  \begin{tabular}{ccccccc}
    \hline
    & Sample & $Log(\tau_V^0)$ & $\alpha$ & $\beta$ & $\gamma$ & Scatter (dex) \\
    \hline
    $\tau_V^F(EL)$ & Discs  & $-4.84$ & $0.39$ & $0.50$ & $-0.99$ & $0.2$ \\
                   & Bulges & $-3.63$ & $0.43$ & $0.34$ & $-0.65$ & $0.2$ \\
    \hline
    $\tau_V^F(CL)$ & Discs  & $-5.59$ & $0.27$ & $0.56$ & $-1.05$ & $0.1$ \\
                   & Bulges & $-6.47$ & $0.21$ & $0.65$ & $-1.03$ & $0.5$ \\
    \hline
    $\tau_V^F(ML)$ & Disc-Dom.  & $-5.50$ & $0.19$ & $0.55$ & $-0.98$ & $0.1$ \\
                   & Bulge-Dom. & $-5.77$ & $0.32$ & $0.58$ & $-0.79$ & $0.4$ \\
    \hline
  \end{tabular}
  \caption{Best-fit parameters for eq.\ref{fit2}, for the different
    subsamples. For the CL and ML libraries the fits are for the
    combined low-z and high-z samples.}
  \label{tab:bfv}
\end{center}
\end{table*}

When we include the effect of complex geometry in the ML, both
$\tau_V^F(EL)$ and $\tau_V^F(CL)$ do not correctly reproduce
$\tau_V^{GS}$. The CL fit parameters still provide a good
approximation for the intrinsic optical depth for the bulge-dominated
subsamples. However, the presence of a bulge perturbs the SED of the
disc-dominated objects. The effect grows when massive bulges are
considered, and it is larger if the bulge component is actively
forming stars. These effects are more important for the high-z
galaxies, where both conditions are frequent in {\gal}. Again we
re-calibrate the best fit parameters for eq.~\ref{fit2} on the
combined (low-z + high-z) ML, and we collect our results in
tab.~\ref{tab:bfv}. We then compare $\tau_V^F(ML)$ and $\tau_V^{GS}$
in fig.~\ref{taufitfig} (right panel). The figure shows that we obtain
a reasonable fit to the combined sample, with a scatter of 0.1 dex and
0.4 dex for the disc and bulge samples respectively. However several
problems arise when we consider the separate samples. We first discuss
the disc-dominated subsamples, where we find an overestimate at the
level of $\sim 1$ dex for the high-z sample. This is due to objects
with a significant bulge component (in mass, star-formation activity
or both). On the other hand, when applied to the bulge-dominated
subsamples, $\tau_V^F(ML)$ provides a systematic overestimate of
$\tau_V^{GS}$ for the high-z sample and a systematic underestimate for
the low-z sample, both at the level of 0.5 dex.

The fitting formula we propose in eq.~\ref{fit2} is analogous to
standard analytic prescriptions, for example the GRV87 and DLB07
approaches. However, it is interesting that the best-fit values for
the parameters $\delta$, $\epsilon$ and $\eta$ differ from the
equivalent values generally used in the literature. Not too
surprisingly, the combination of best-fit parameters for $M_{\rm gas}$
and $R_{\rm gal}$ hints of a dependence of $\tau_V^{GS}$ on the cold
gas surface density (for all fits in tab.~\ref{tab:bfv} $\gamma \sim 2
\beta$). This is similar to eq.~\ref{dlb}, even if in the latter
equation the dependence is linear with the cold gas surface density.
The dependence of $\tau_V^{GS}$ on metallicity $Z$ is completely
different, and much weaker than in eq.~\ref{grv}.


\section{Comparisons at Optical Wavelengths}
\label{section:compattenu}

\begin{figure*}
  \centerline{
    \includegraphics[width=9cm]{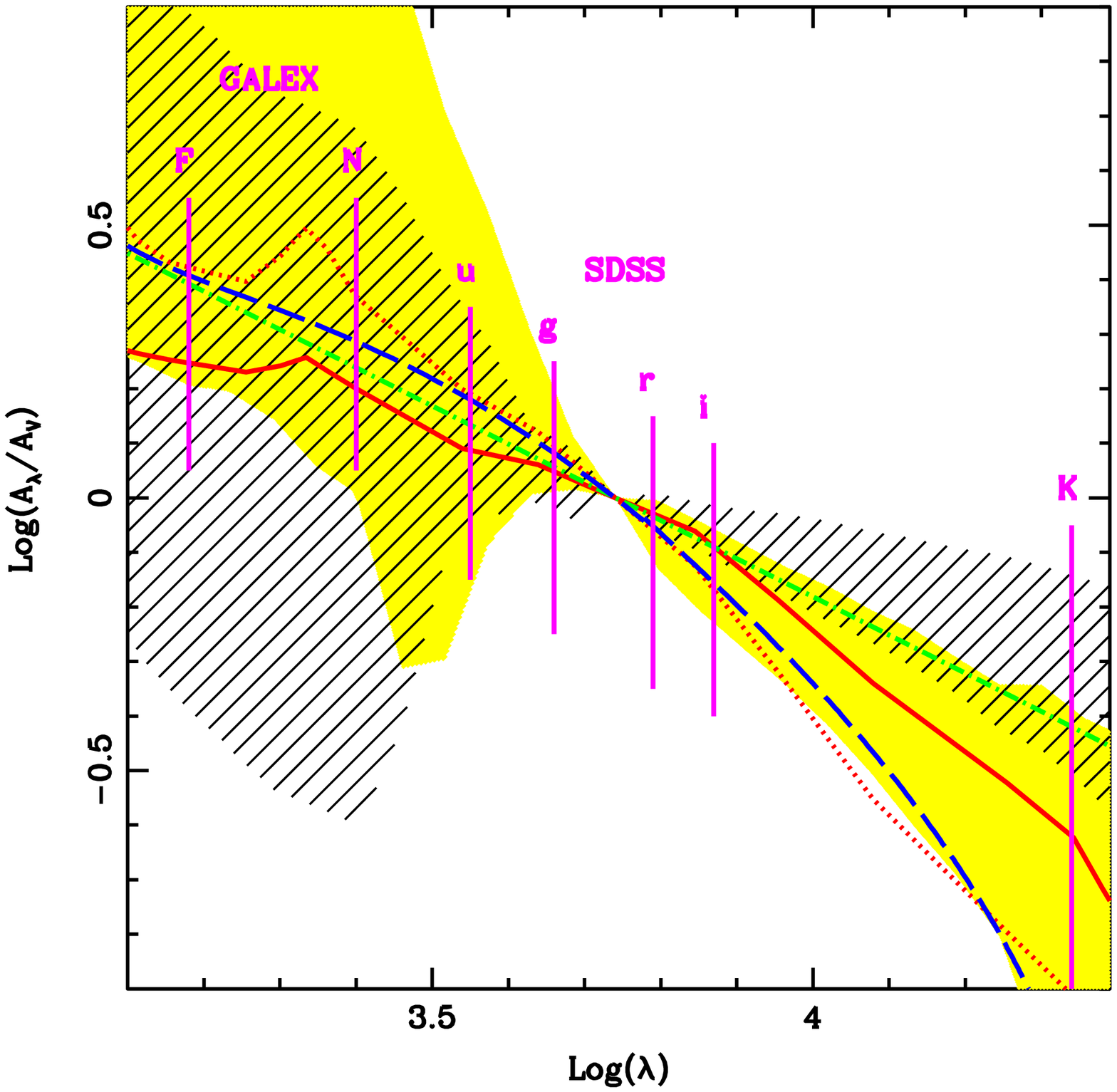}
    \includegraphics[width=9cm]{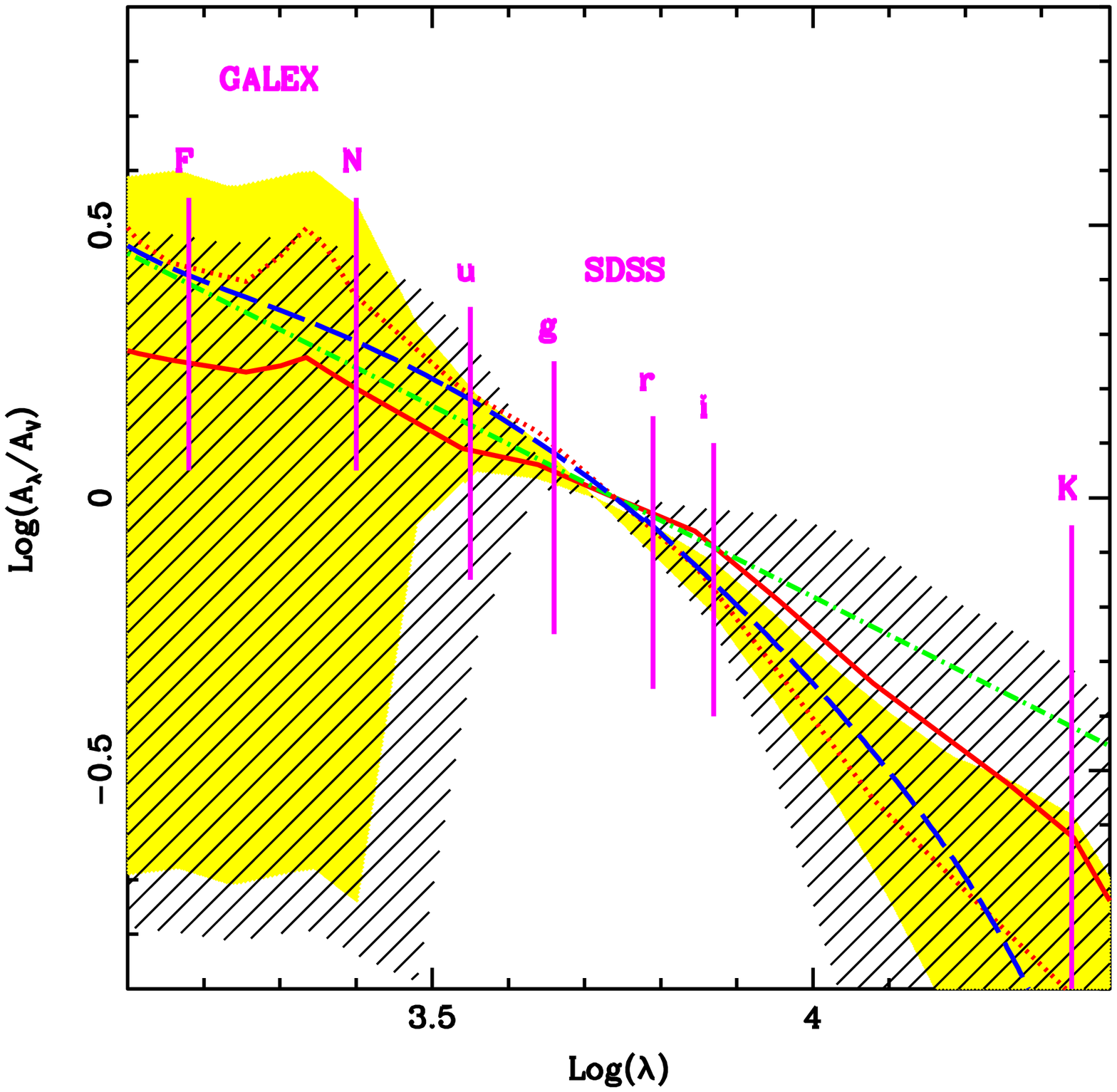}
  }
  \caption{Left panel: The V-band normalized attenuation curves
    obtained by {\gal}+{\gs} for the low-z and high-z samples (yellow
    shaded area and diagonal texture respectively), are shown compared
    with the MW extinction law (dotted red line, by Mathis et
    al. 1983), the Calzetti et al. (dashed blue line), the CF00
    (dot-dashed green line), and the {\it slab} + with the MW
    extinction (continuous red line) attenuation curves. Right panel:
    the same quantities for Prescription TAU-GS (see
    sect.~\ref{prescriptions}).}
    \label{gs_extlaw}
\end{figure*}

\begin{table}
\begin{center}
  \begin{tabular}{ll}
    \hline
    Prescription & Optical Depth  \\
    \hline
    TAU-GS & ``true'' optical depth from {\gs} ($\tau_V^{GS}$)  \\
    TAU-FIT & results of fit to {\gs} libraries ($\tau_V^F(ML)$)  \\
    TAU-GRV & Guiderdoni \& Rocca-Volmerange (1987; GRV87) \\
    TAU-DLB & De Lucia \& Blaizot, (2007; DLB07)  \\
    \hline
  \end{tabular}
  \caption{Dust Prescriptions}
  \label{tab:models}
\end{center}
\end{table}

In this section, we compare the results of the {\gs} libraries with
several approximations. In all of these approximations, we assume that
the wavelength and geometry dependent behavior of the dust extinction
can be captured by using an expression that relates the effective
optical depth $\tau_V$ to physical properties of the galaxy, in
combination with a ``slab'' model for the geometry and the DLB07
composite MW + CF00 attenuation curve (see \S\ref{section:ricette};
following DLB07, we select $\mu$ randomly from a Gaussian
distribution).  We assume a single dust slab to describe the
attenuation of both disc and bulge. To assure proper comparison with
the {\gs} SEDs, we also average slab model results over inclination.
We will refer to the four prescriptions that we will test:

\begin{itemize}
\item {\bf Prescription TAU-GS}: we use the actual value of the
  effective optical depth $\tau_V^{GS}$ from the RT library.

\item {\bf Prescription TAU-FIT}: we use the fitting formula
  $\tau_V^F(ML)$ derived in \S~\ref{sec:physical}.

\item {\bf Prescription TAU-GRV}: we use the GRV87 formula
(Eqn.~\ref{grv}) for $\tau_V$. Recall that in this approximation, the
column density of the gas is simply proportional to the gas fraction
(no scaling with galaxy radius).

\item {\bf Prescription TAU-DLB}: we use the DLB07 formula
(Eqn.~\ref{dlb}) for $\tau_V$. Here, the optical depth is proportional
to the surface density of the gas (scales as $1/r_{\rm gal}^2$).

\end{itemize}
These prescriptions are summarized in Table~\ref{tab:models}.

We have extensively tested many different possible combinations of
ingredients, for example, substituting the Calzetti or CF00
attenuation law for the Galactic extinction law, or substituting the
oblate ellipsoid geometry for the slab model. We find a similar level
of agreement with the Calzetti law, but worse agreement with the CF00
attenuation law. The oblate ellipsoid geometry produces similar
results to the slab model.

\subsection{Attenuation curves}
\label{section:attcurve}

In Fig.~\ref{gs_extlaw} (left panel) we show the V-band normalized
attenuation curves $A_\lambda^{GS}$ for the {\gs} SEDs in the ML. The
yellow shaded area is for the low-z sample, the diagonal texture for
the high-z sample (1--$\sigma$ confidence regions). It is worth noting
that the galaxy-to-galaxy scatter shown in fig.~\ref{gs_extlaw} is an
underestimate, as we do not consider the distribution of inclination
angles (but only angle-averaged SEDs) and we keep fixed both $t_{\rm
  esc}$ and the intrinsic properties of dust grains. For comparison we
also display the average MW extinction curve (by Mathis et al.  1983,
dotted red line), and the attenuation curves of Calzetti et al.
(2000) (dashed blue line), CF00 (dot-dashed green line), and the slab
+ MW extinction + scattering (continuous red line). All curves have
been normalized to $\tau_V=1$ in order to highlight the different
dependencies on the wavelength.

The comparison with the Calzetti et al curve is particularly
interesting, since its shape has been interpreted in several ways. For
instance, Gordon, Calzetti, \& Witt (1997) ascribed the observed shape
to the presence of dust lacking the $2175$ \AA\ feature in the
extinction curve, i.e. intrinsic dust differences. This interpretation
is probably a consequence of adopting a model that accounts for the
clumping of dust but not of stars, and with a spatial distribution of
stars independent of stellar age. Granato et al. (2000) reproduced
this featureless and shallow attenuation as a result of the complex
and wavelength-dependent geometry where the UV-emitting stars are
heavily embedded inside MCs, while older stars, mainly emitting in the
optical and NIR, suffer a smaller effect from the diffuse medium.
Similar conclusions on the strong effect of complex geometries in the
resulting attenuation have been found in the analysis by e.g. Panuzzo
et al. (2003, 2007), Pierini et al. (2004), Tuffs et al. (2004), Inoue
(2005), and Inoue et al. (2006) . While one can be confident that
intrinsic dust differences are effectively present in different
galaxies, as well as in different environments, it also appears that
the complex geometry in galaxies can have a major and sometimes
dominant role.

Since, as described in \S\ref{section:grasil}, in the {\gal}+{\gs}
models the dust properties (size distribution and composition) are
fixed so as to reproduce the average MW {\it extinction} curve, any
difference between that and the {\it attenuation} curves we obtain,
must be ascribed to the particular geometrical configuration of each
model galaxy. As we said, the geometry has to be understood in a broad
sense, i.e. not only the stars and dust distributions in a bulge
and/or in a disc, but also the clumping of stars and dust and the
age-, therefore wavelength-dependence of the amount of attenuation
suffered by stars. We stress that the attenuation curve we obtain for
each model would be exactly equal to the adopted extinction curve if
we were to assume that all the dust is in a foreground screen in front
of the stars and no scattered light reached the observer. Therefore
the very large spread we get, as evident from the figure, has to be
ascribed to the accounting of the star formation histories, the
distribution of stars and dust with the appropriate radii and the
connection between stellar populations of different ages with
different dusty environments.

We also show in fig.~\ref{gs_extlaw} (right panel) the corresponding
mean attenuation as a function of wavelength for Prescription TAU-GS;
i.e., using the composite attenuation law suggested by DLB07 based on
the CF00 model.  Encouragingly, the behavior is qualitatively similar
to that of the the intrinsic {\gs} mean attenuation curve. The major
discrepancies arise at very short wavelengths (in both samples) where
{\gs} predicts larger attenuations, and in the near-infrared region
(in the high-z sample), where {\gs} predicts a smaller scatter and on
average larger attenuations. We test that the scatter in the
attenuation laws at short wavelengths is reduced if we assume a fixed
value for the $\mu$ parameter (i.e. $\mu=0.3$) in Prescription TAU-GS,
instead of using a Gaussian distribution.

\subsection{Attenuation and color excess}
\label{prescriptions}

In this sub-section we show the results we obtain for attenuation and
reddening by coupling {\gal} either with the full {\gs} RT
calculation, or by applying different prescriptions for attenuation to
the pure stellar SED (summarized in Table \ref{tab:models}). We
illustrate our comparison by showing attenuations and color excesses
as a function of $M_\star$ and specific star formation rate ${\rm
  SFR}/M_\star$. We show the results for the $FUV$ GALEX filter, the
$g$ and $r$ SDSS filters and for the $K$-band bandpass, but similar
results holds if we consider the full set of SDSS bandpasses plus
near-infrared photometry. We also consider bulge- and disc-dominated
subsamples separately. We consider restframe magnitudes both for the
low-z and high-z samples. This choice allow us to test the robustness
of different prescription for dust attenuation, when changing the
properties of the underlying galactic populations.

\begin{figure*}
  \centerline{
    \includegraphics[width=9cm]{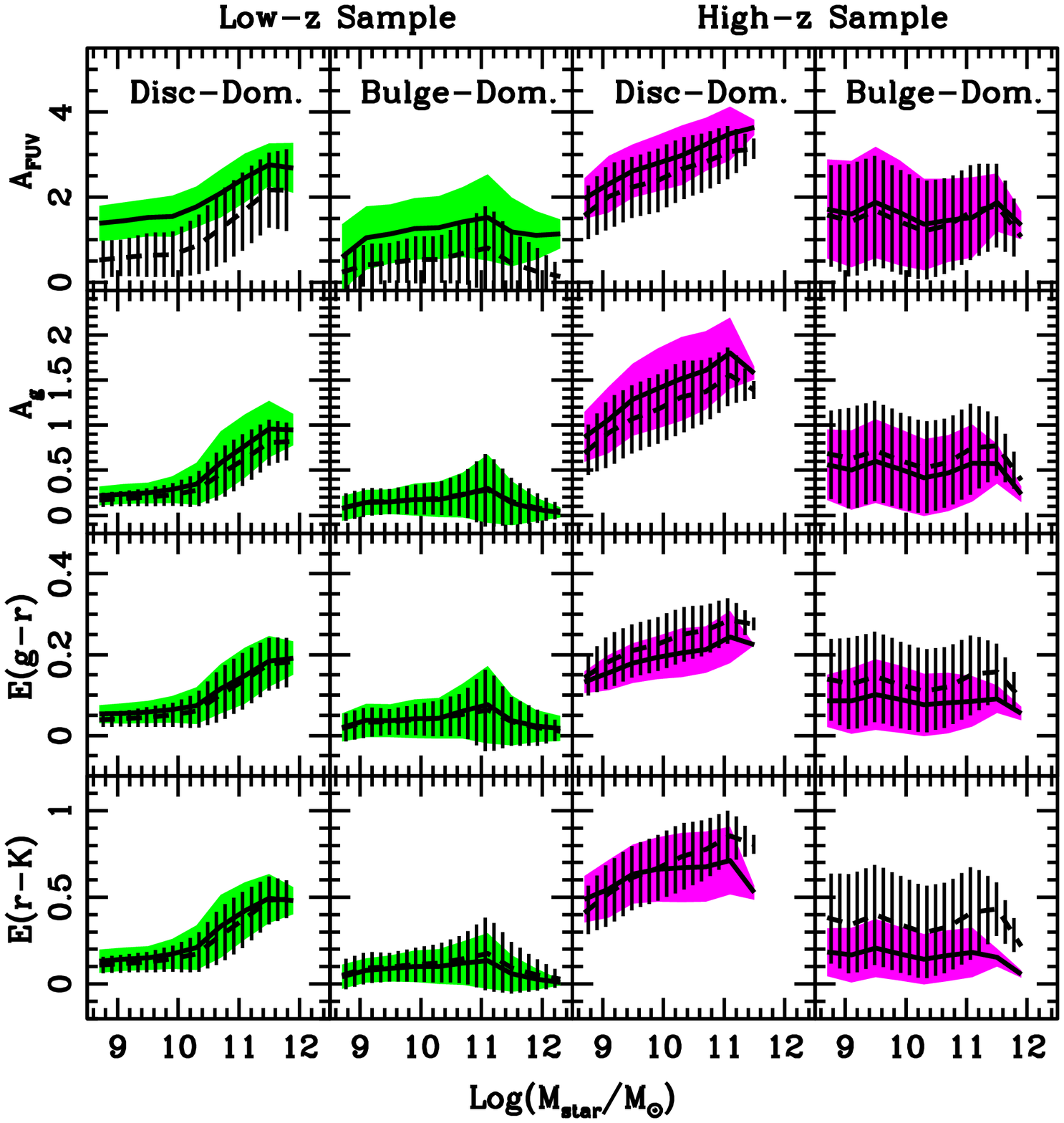}
    \includegraphics[width=9cm]{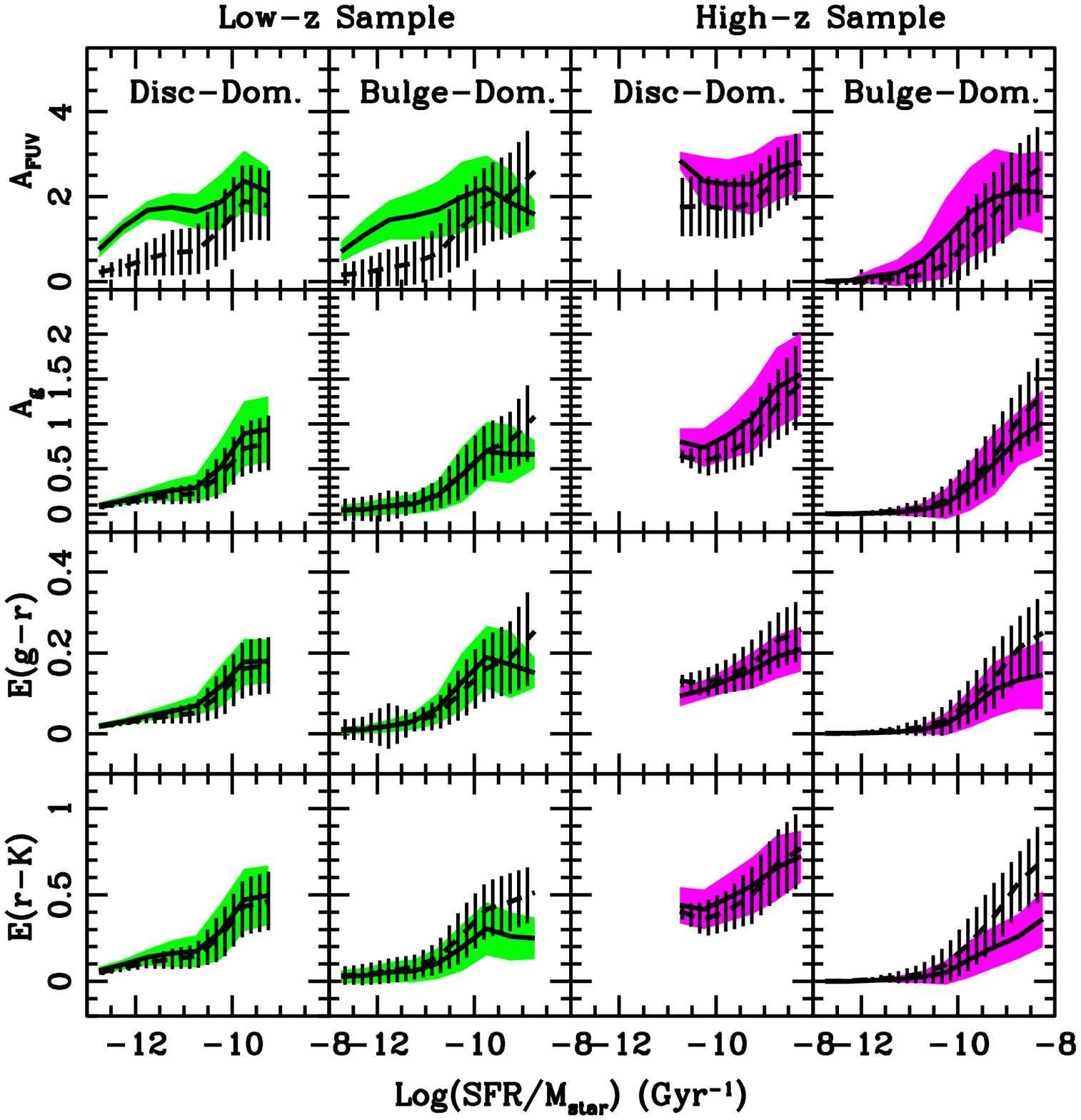}
  }
  \caption{Comparison between the color excesses $E^{GS}(g-r)$ and
    $E^{GS}(r-K)$, and the attenuations $A^{GS}_{FUV}$ and $A^{GS}_g$
    (shaded areas)as a function of stellar mass $M_\star$, (left
    panel) and of specific star formation rate (right panel), with the
    predictions of Prescription TAU-GS (vertical texture). We show the
    results separately for the disc- and bulge-dominated samples and
    for the low-z and high-z samples, as labeled on the figure. }
  \label{mor15}
\end{figure*}

In Fig.~\ref{mor15} we compare $E^{GS}(g-r)$, $E^{GS}(r-K)$,
$A^{GS}_{FUV}$ and $A^{GS}_g$ to the corresponding quantities obtained
when $\tau_V^{GS}$ is combined with a composite attenuation law like
DLB07 (Prescription TAU-GS; shaded areas and diagonal texture
represent the 1--$\sigma$ confidence regions). Clearly, this case
provides a reference point for whole analytic approach --- in
principle, none of our other prescriptions can possibly do better at
reproducing the full {\gs} results than this one, where the effective
optical depth is measured directly from the RT libraries.

Indeed, we see that this model provides a satisfactory representation
of the RT results at optical wavelengths. Mean attenuations are
slightly underpredicted, but the trends as a function of $M_\star$ and
$SSFR$ are correctly recovered. However the agreement worsens at
shorter wavelengths, with a significant underprediction of the level
of attenuation. This result is expected: at FUV wavelenghts the
attenuation is extremely sensitive to the details of the dust
distribution and composition, and it is very difficult to describe it
with a simple analytic formulation.  This result can also be
understood in terms of the different distributions of attenuation that
we have already seen in Fig.~\ref{gs_extlaw}. The same prescription is
able to reproduce $A^{GS}_{FUV}$ and $A^{GS}_g$ for the high-z sample,
but the agreement is poor for both color excesses, particularly in the
disc-dominated sub-sample.  Again, this reflects the disagreement in
the attenuation curves.

\begin{figure*}
  \centerline{
    \includegraphics[width=9cm]{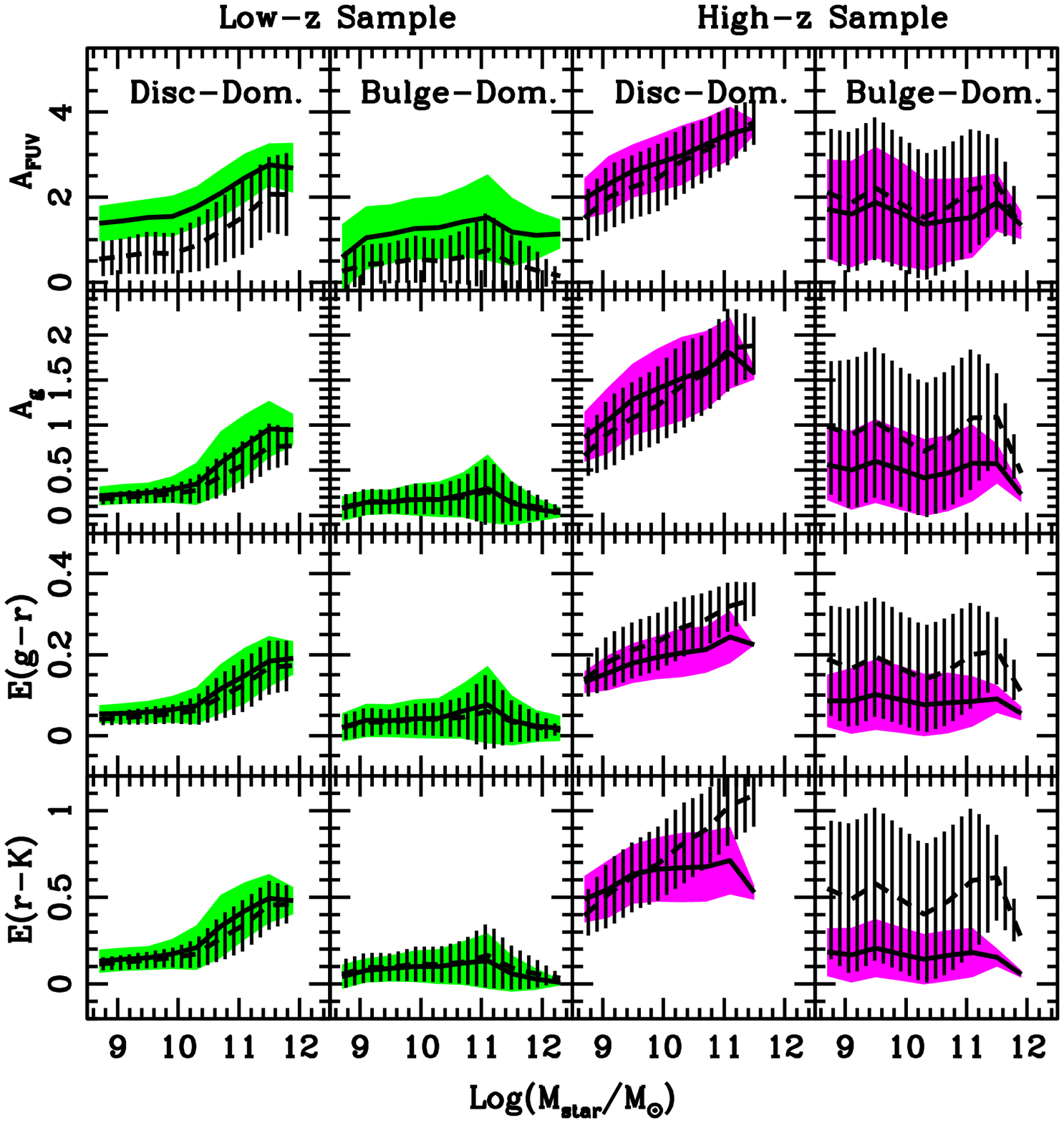}
    \includegraphics[width=9cm]{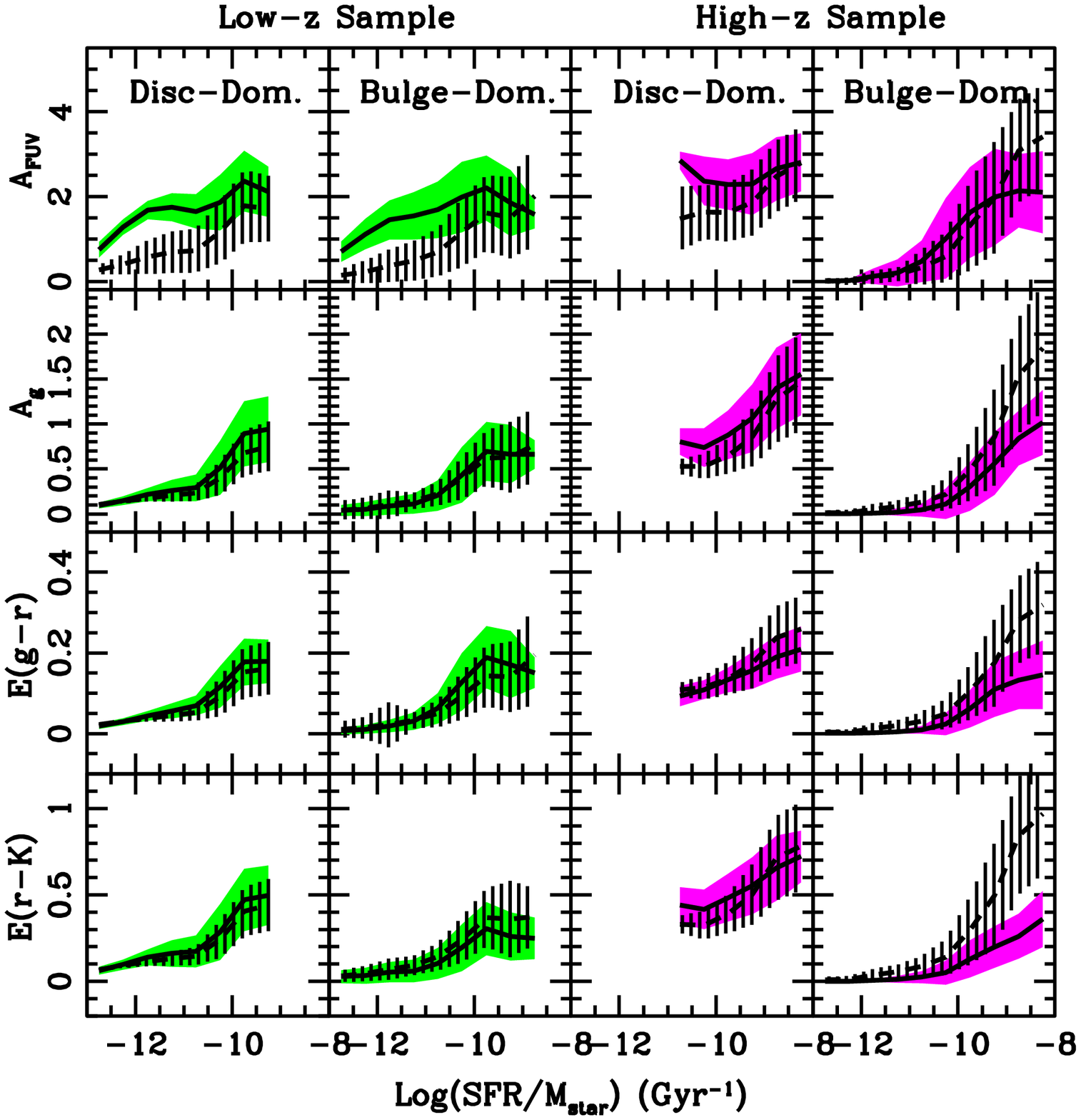}}
  \caption{Same as in fig.~\ref{mor15}. The vertical textured area
    shows the predictions of Prescription TAU-FIT.}
  \label{mor20}
\end{figure*}

We compare the same quantities for our TAU-FIT prescription in
fig.~\ref{mor20}. The agreement is again satisfactory, and more
importantly where there are discrepancies, they are similar to those
seen in Prescription TAU-GS. This indicates that the main source of
remaining error is not our fitting results for $\tau_V$ but rather the
simplified approach we have used to estimate the attenuation curve.

\begin{figure*}
  \centerline{
    \includegraphics[width=9cm]{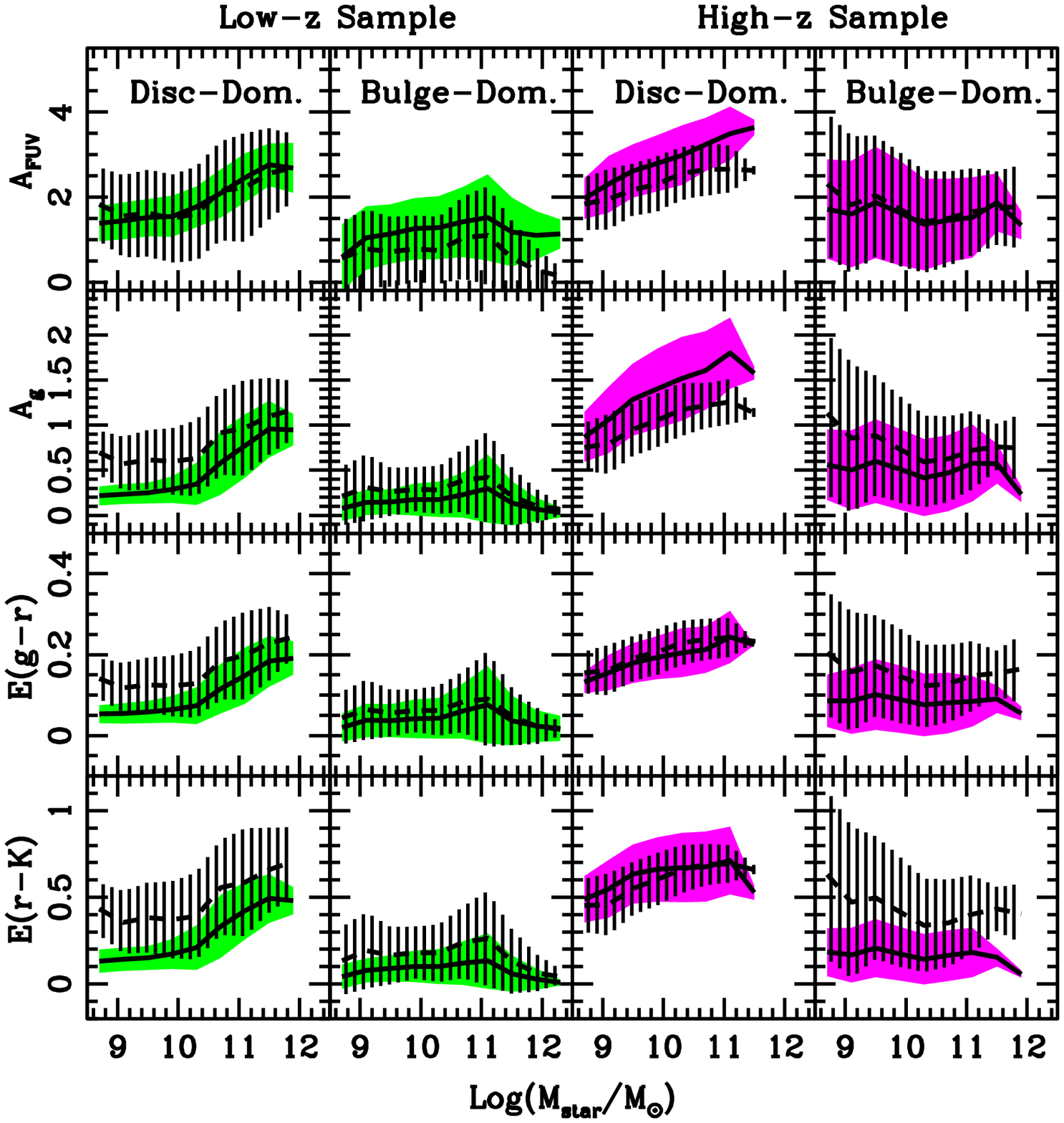}
    \includegraphics[width=9cm]{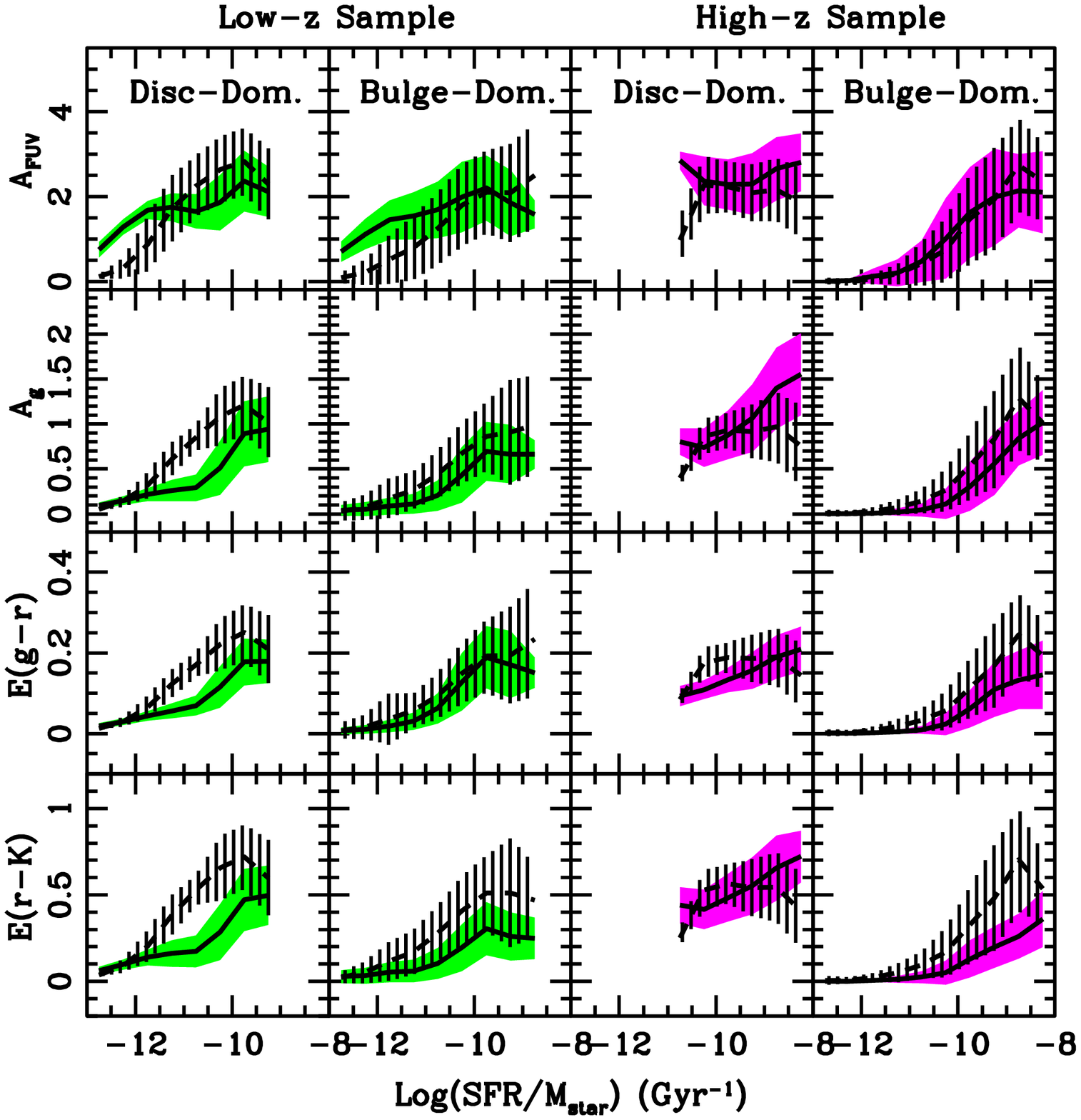}}
  \caption{Same as in fig.~\ref{mor15}. The vertical textured area
    shows the predictions of Prescription TAU-GRV.}
  \label{mor22}
\end{figure*}
\begin{figure*}
  \centerline{
    \includegraphics[width=9cm]{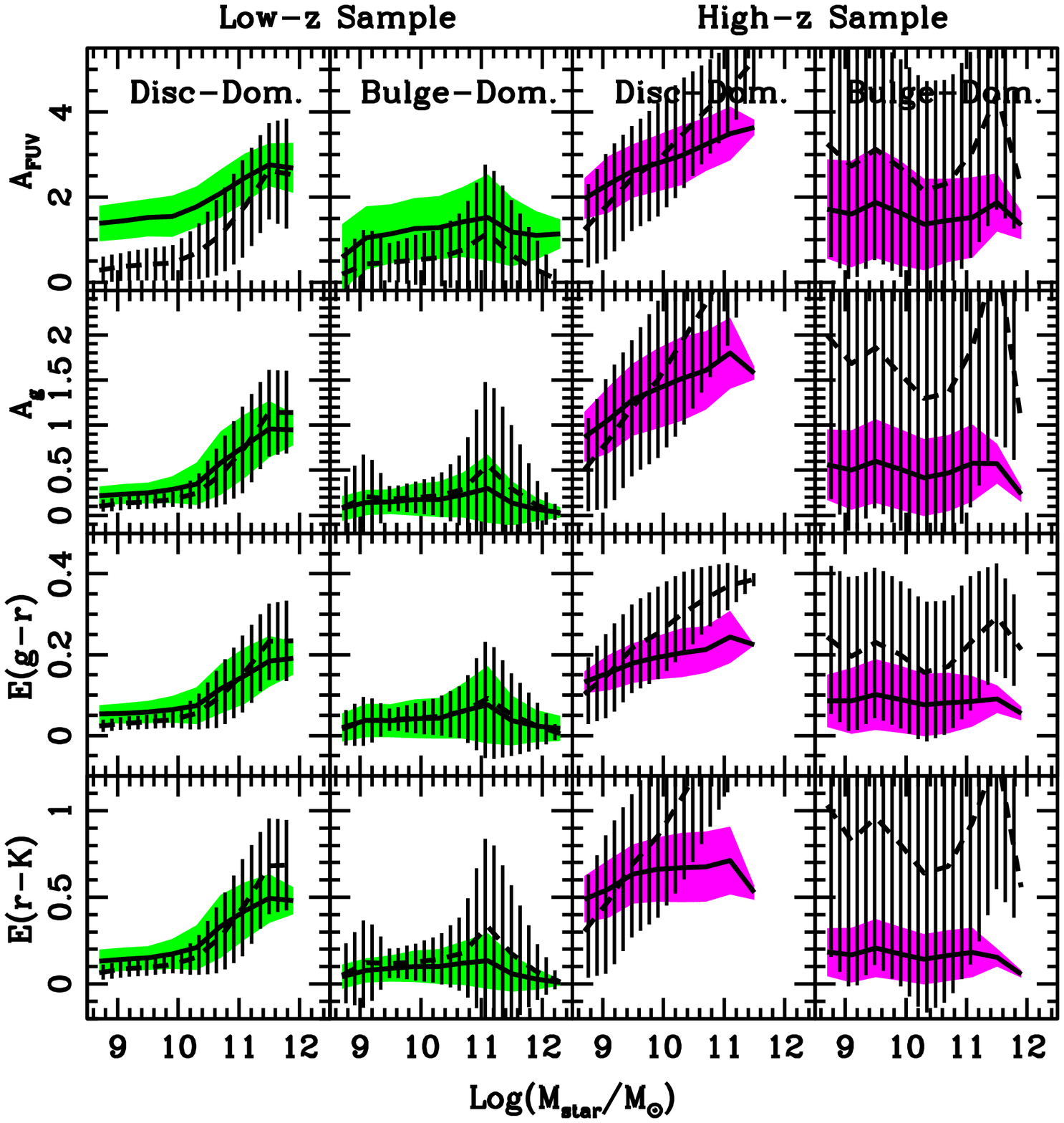}
    \includegraphics[width=9cm]{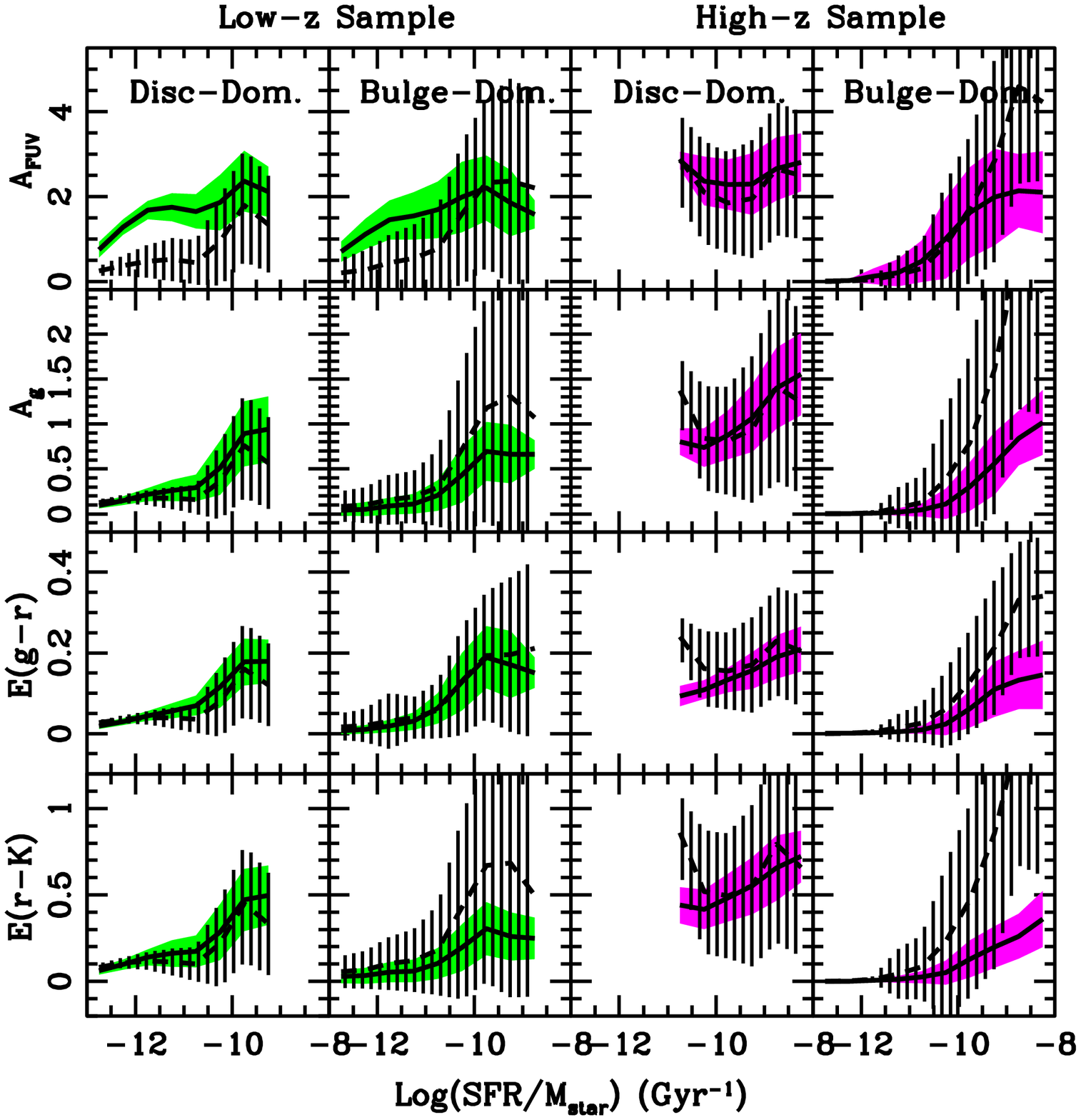}}
  \caption{Same as in fig.~\ref{mor15}. The vertical textured area
    shows the predictions of Prescription TAU-DLB.}
  \label{mor24}
\end{figure*}
We show in fig.~\ref{mor22} the same comparison with Prescription
TAU-GRV. In the low-z sample (left panel), the ranges covered by the
model and the prescription are in reasonable agreement, despite a
slight systematic overestimate of attenuation and color excess in both
the disc- and in the bulge-dominated sample. The dependence of
attenuation on $M_\star$ is at least approximately reproduced. The
dependence of attenuation and color excess on $M_\star$ are only
marginally reproduced, with a larger discrepancy for the
bulge-dominated objects. Moreover, the distributions of attenuations
and color excesses around the mean values show a smaller (larger)
scatter with respect to {\gs} predictions for disc-(bulge-)dominated
objects. The effect of considering a simple geometry is similar to the
low-z case (higher attenuations and color excesses for the pure discs
and vice-versa for the pure bulges).

Finally, we show in fig.~\ref{mor24} the comparison with Prescription
TAU-DLB. Following the original work, for each object we compute
$r_{1/2}$ starting from the disc scale-radius (even if the galaxy is
bulge-dominated). Prescription TAU-DLB is able to reproduce both {\gs}
attenuations and color excesses and their dependence on $M_\star$ in
the low-z sample, with the notable exception of the far-UV
attenuations. Moreover, the scatter is always larger than the
corresponding quantity in {\gs}.  Quite interestingly a very similar
behavior is obtained for the CL, with only small differences with
respect to the ML. We then conclude that Prescription TAU-DLB is less
sensitive to the composite vs simple geometry than Prescription
TAU-GRV. On the other hand, when we compare Prescription TAU-DLB to
the high-z sample we obtain both a steeper dependence of attenuation
from $M_\star$ and a larger scatter with respect to {\gs}.

Both Prescriptions TAU-GRV and TAU-DLB provide acceptable agreement
with the {\gs} results for the low-z samples, but they do not do well
at high redshift. The satisfactory agreement at low-z is not
surprising, given the fact that the original prescriptions are
calibrated to reproduce the properties of galaxies in the local
Universe. The origin of the high-z discrepancy lies in the different
distribution of cold gas fractions and surface densities in the two
samples (the dependence on $Z$ and $M_{\rm gas}$ being the same). The
DLB07 prescription for $\tau_V$ is proportional to the gas surface
density; we demonstrate that this assumption is broadly consistent
with the RT results, however our analysis suggests a much weaker
dependence of $\tau_V$ on surface density (to the square-root power).
We therefore expect the differences to grow at high-z, where gas
surface densities are considerably higher.

\begin{figure*}
  \centerline{
    \includegraphics[width=15cm]{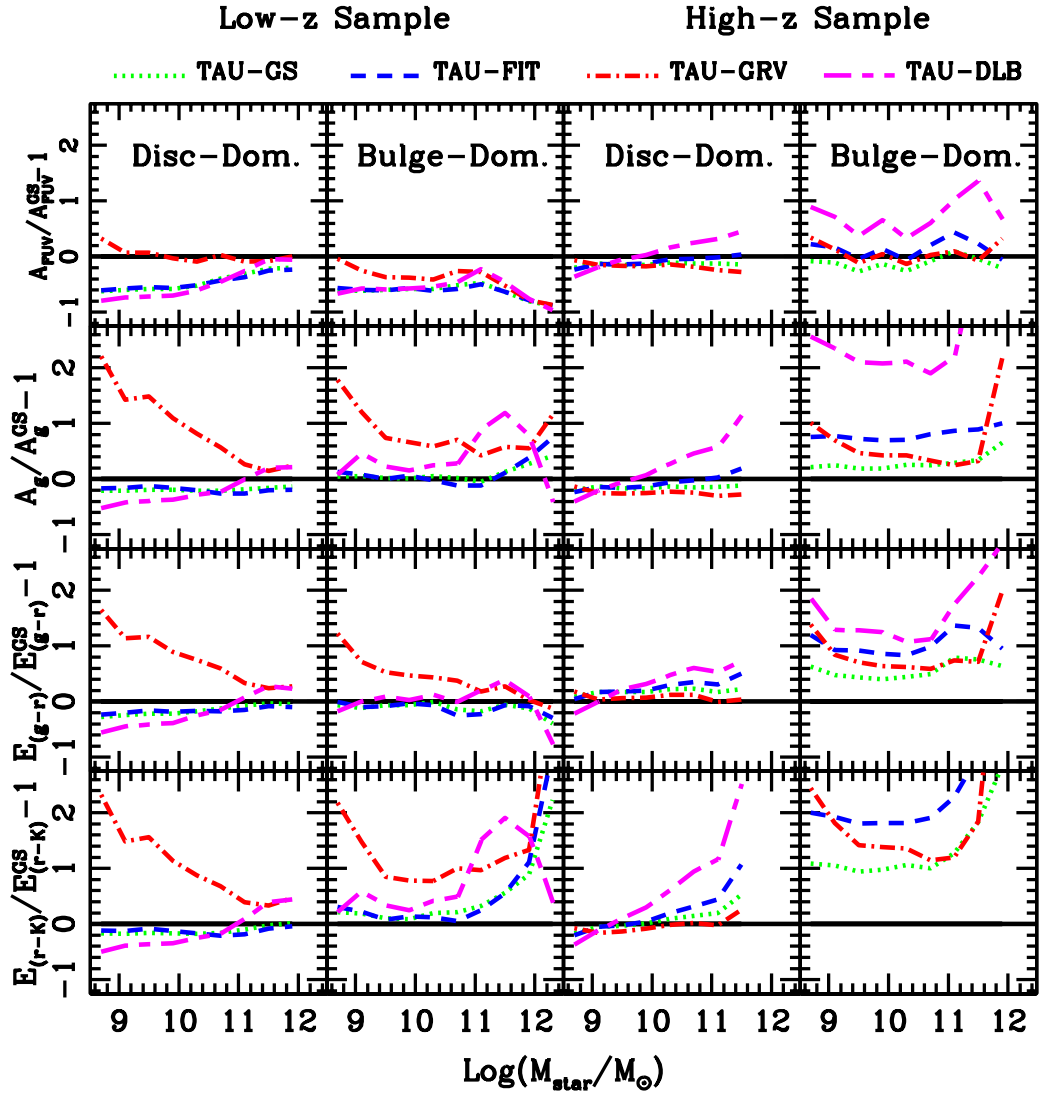}
  }
  \caption{Residuals in the mean attenuations as a function of stellar
    mass for our four prescriptions with respect to {\gs} predictions.
    Dotted, dashed, dot-dashed, long-short dashed lines refer to
    prescription TAU-GS, TAU-FIT, TAU-GRV and TAU-DLB respectively.}
  \label{mean_res}
\end{figure*}
\begin{figure*}
  \centerline{
    \includegraphics[width=15cm]{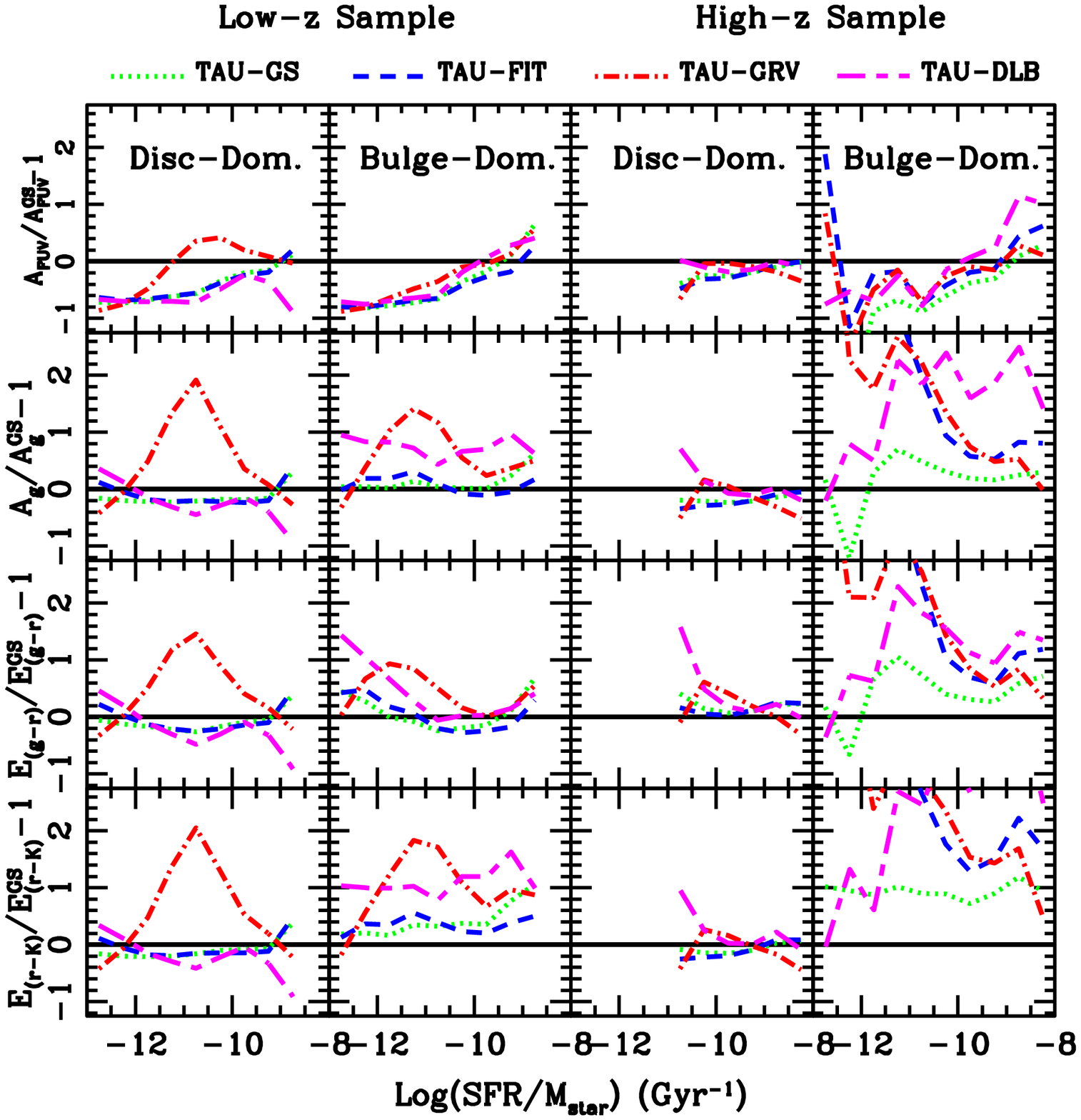}
  }
  \caption{Residuals in the mean attenuations as a function of
    specific star formation rate for our four prescriptions with
    respect to {\gs} predictions. Lines as in fig.~\ref{mean_res}.}
  \label{mean_res_b}
\end{figure*}
\begin{figure*}
  \centerline{
    \includegraphics[width=15cm]{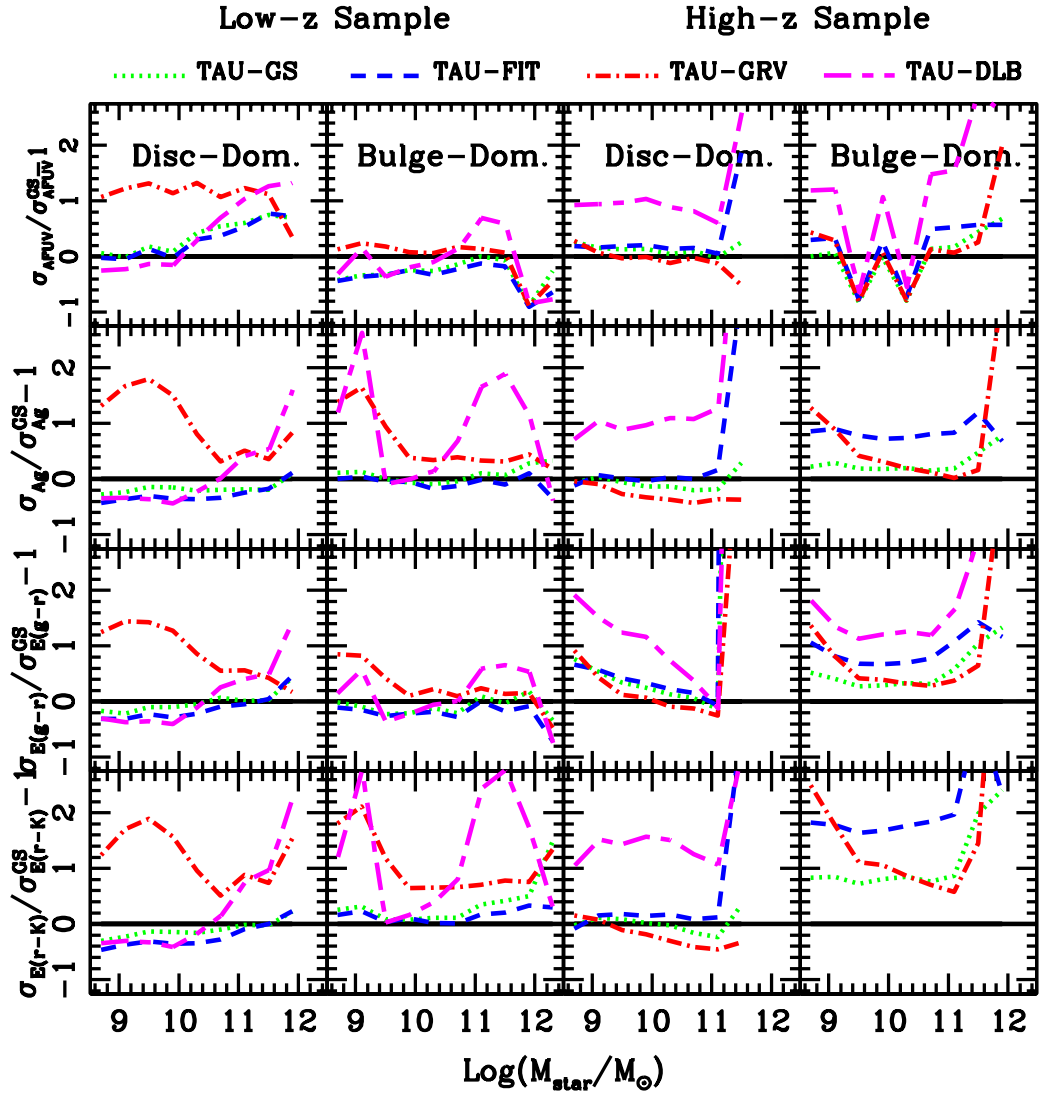}
  }
  \caption{Residuals in the variance in the mean attenuations as a
    function of stellar mass for our four prescriptions with respect
    to {\gs} predictions. Lines as in fig.~\ref{mean_res}.}
  \label{var_res}
\end{figure*}
\begin{figure*}
  \centerline{
    \includegraphics[width=15cm]{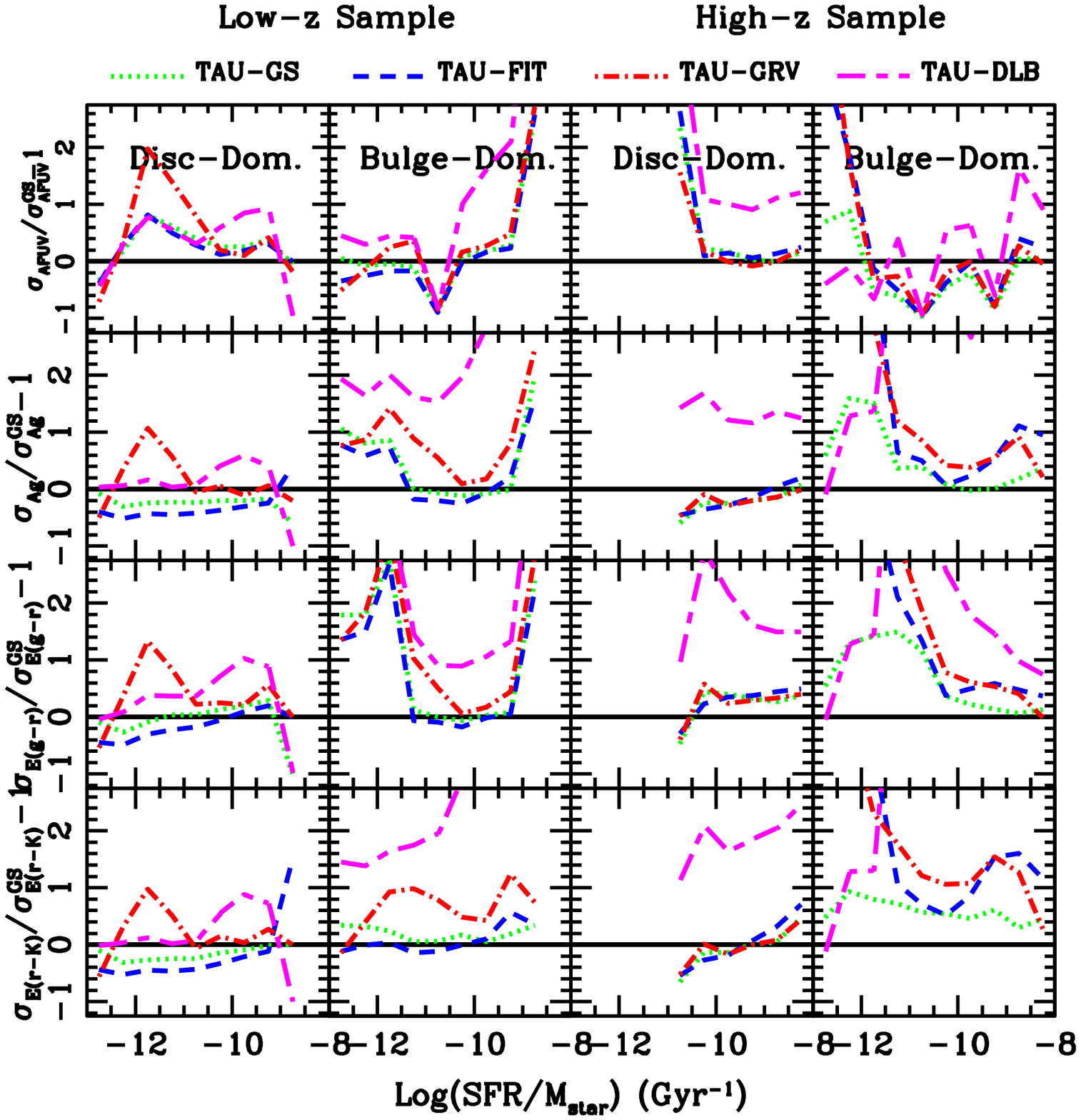}
  }
  \caption{Residuals in the variance in the mean attenuations as a
    function of specific star formation rate for our four
    prescriptions with respect to {\gs} predictions. Lines as in
    fig.~\ref{mean_res}.}
  \label{var_res_b}
\end{figure*}

From fig.~\ref{mean_res} to~\ref{var_res_b} we summarize the results
for all four prescriptions in terms of the residuals of the difference
between the mean attenuation predicted by the prescription and the
intrinsic attenuation in {\gs}. From these figures, we can clearly see
that prescription TAU-GS always has the smallest residuals, as
expected. Prescription TAU-FIT is seen to be quite successful in that
in most cases, the residuals are similar to those of prescription
TAU-GS and the trends are similar. Prescriptions TAU-GRV and TAU-DLB
show a similar level of agreement overall, but often show different
trends with galaxy properties such as stellar mass. All prescriptions
seem to have particular difficulty reproducing the {\gs} results for
high-z bulge-dominated galaxies.

\begin{figure*}
  \centerline{
    \includegraphics[width=15cm]{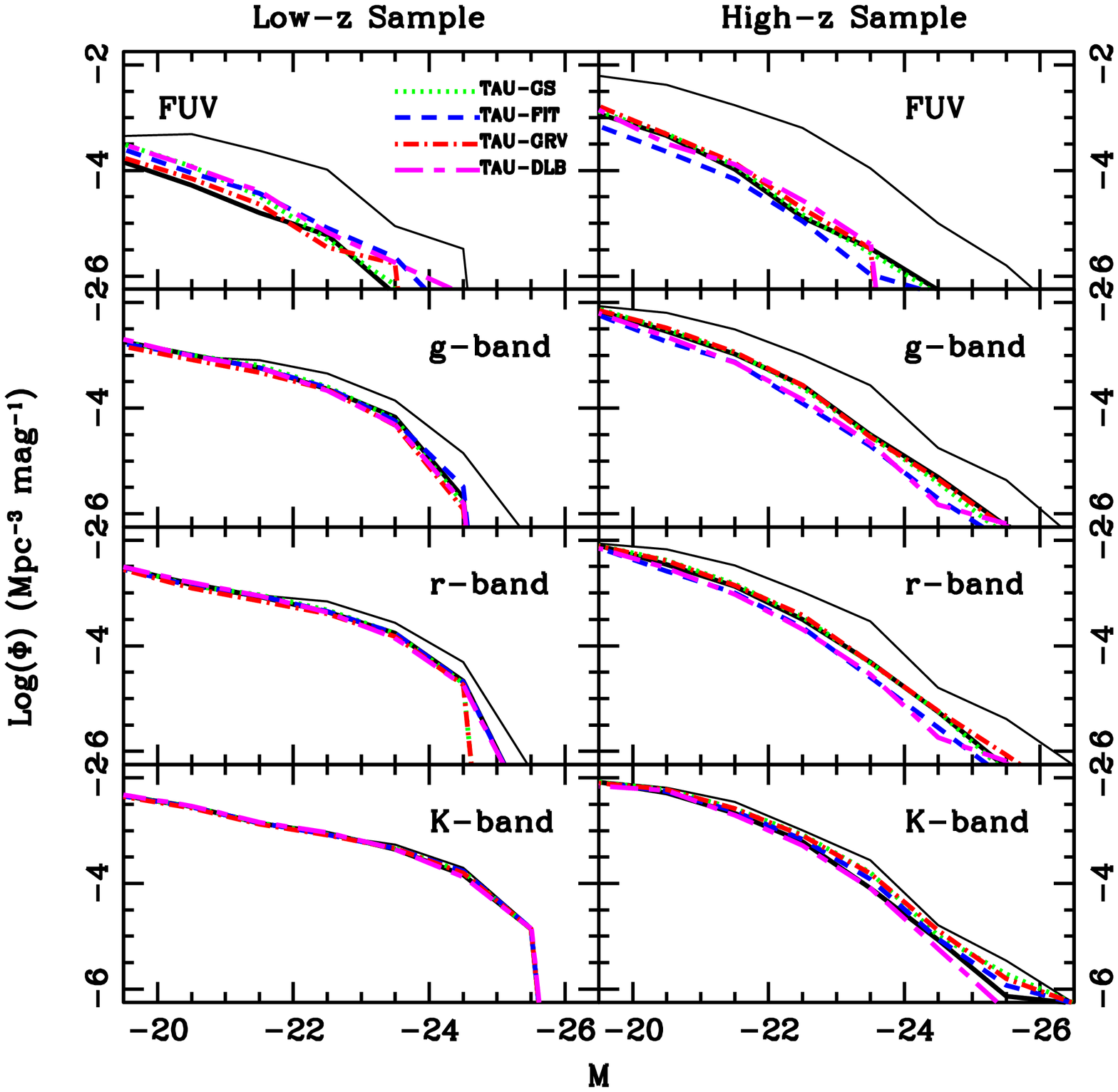}
  }
  \caption{Effect of the different prescriptions for dust absorption
    on the galaxy luminosity function in different bands. Lines as in
    fig.~\ref{mean_res}, thin solid line refers to the unextinguished
    luminosity function.}
  \label{lfest}
\end{figure*}

Finally, we assess the impact of the different dust prescriptions on
the predictions of the statistical properties of galactic samples. In
particular, in fig.~\ref{lfest} we focus on the resulting luminosity
functions, which are commonly used as a baseline prediction for
semi-analytic models. The thin solid line represents the
unextinguished luminosity function at different wavelengths for the
low-z and high-z samples, whereas the thick solid line shows the
luminosity function of the samples as predicted by {\gs}. The other
lines represent Prescription TAU-GS, TAU-FIT, TAU-GRV and TAU-DLB (the
dotted, dashed, dot-dashed and long-short dashed line respectively).
This comparison is quite encouraging, in that it indicates that, while
it is extremely important to correct for the effects of dust,
particularly at high redshift and in shorter wavelength bands, even a
simple analytic approach gives quite good results, when dealing with
integrated quantities, such as the LFs, in which galaxy to galaxy
differences are smoothed out, particularly in the optical-NIR region
characterized by relatively regular and featureless SEDs. However, the
spread in the predicted LF increases at UV wavelengths. As we have
seen, the effect of dust on galaxy colors is not reproduced as well by
the simple analytic prescriptions.

\subsection{Inclination dependence of attenuation}
\label{inclination}

\begin{figure}
  \centerline{
    \includegraphics[width=9cm]{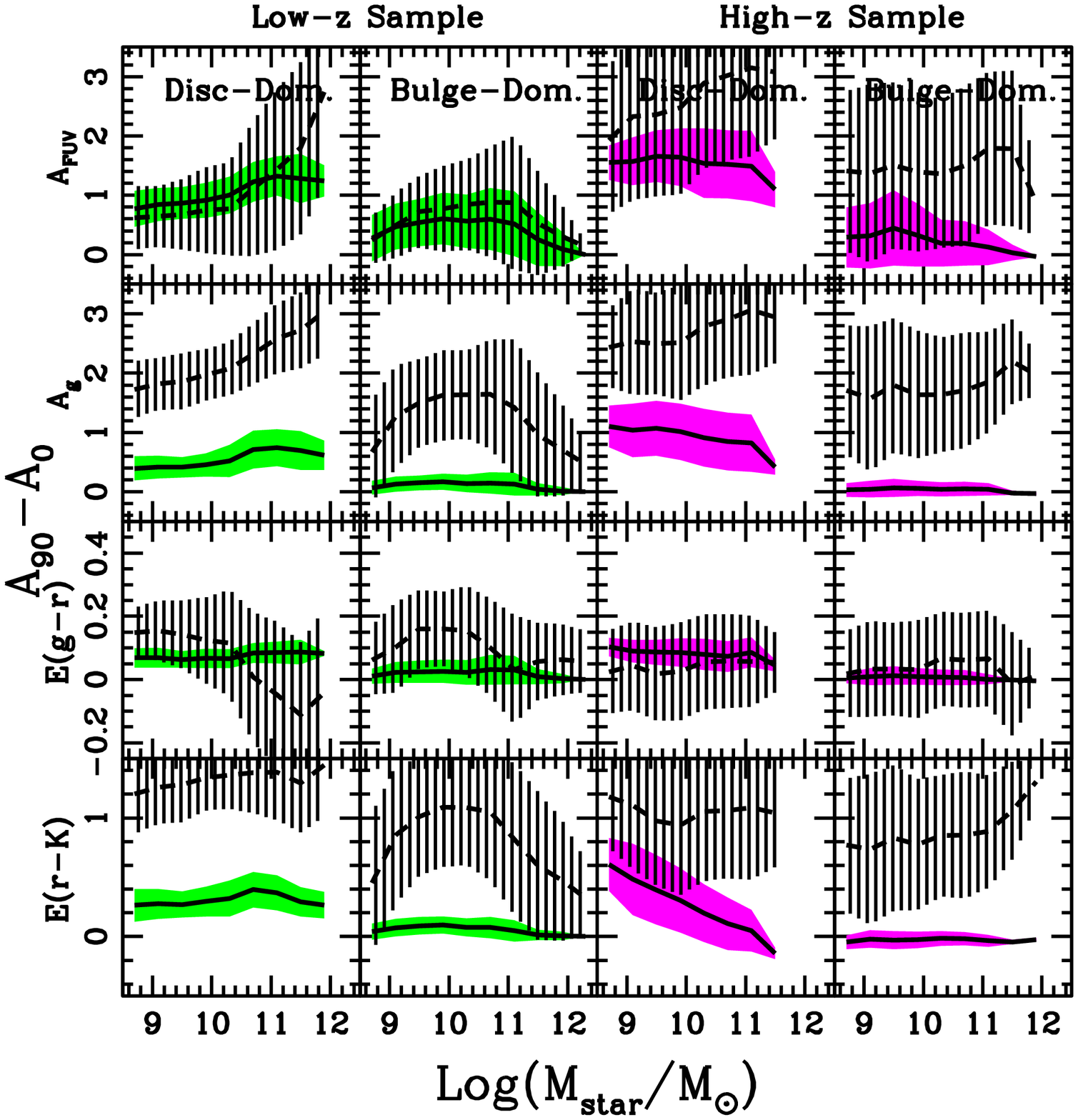}
  }
  \caption{Edge-on relative to face-on attenuation, $A_{90}-A_0$, as a
    function of stellar mass. The shaded area shows the GRASIL
    results, the vertical texture the prediction of Prescription
    TAU-DLB}
  \label{figincl90vs00}
\end{figure}

The geometrical dependence of the predicted SEDs manifests itself also
in the dependence of attenuation on the viewing angle.  In many
papers, statistical estimates of the dust opacity in galaxy discs have
been worked out, through the analysis of the inclination dependence of
apparent magnitudes and colors of large samples of spiral galaxies
(e.g. de Vaucouleurs et al. 1991; Giovanelli et al.  1995; Moriondo,
Giovanelli \& Haynes, 1998; Tully et al. 1998; Graham 2001; Masters,
Giovanelli \& Haynes, 2003). Extensive work on the angle dependence
predicted by RT models (chiefly for disc galaxies) has also been
presented in the literature (e.g. Kylafis \& Bahcall 1987; Byun,
Freeman \& Kylafis 1994; Bianchi, Ferrara \& Giovanardi 1996;
Kuchinski et al. 1998; Tuffs et al. 2004; Pierini et al. 2004; Rocha
et al. 2008). The interpretation of observations to infer opacities,
scale-lengths, luminosities etc. are non-trivial, because of the
radial gradients in the dust distribution and metallicity thus in the
dust opacity, the different scale-heights of stars and dust, the bulge
to disc ratio, and the inclination effects (e.g Xilouris et al.  1999;
Boissier et al. 2004; Popescu et al. 2005; Mollenhoff, Popescu \&
Tuffs 2006; Driver et al. 2007).

The statistical data for the attenuation-inclination dependence may
therefore provide a further test for SAM predictions (see e.g.
Granato et al. 2000), if the spectral properties computed for the SAM
are reliable, given the complexities described above.  In the previous
Sections we have considered the results we obtain for angle-averaged
SEDs. In Fig. \ref{figincl90vs00} we show the dust net attenuation
(difference edge-on to face-on) predicted using GRASIL and
Prescription TAU-DLB, which we recall adopts an infinite plane
parallel {\it slab} (Eq.~\ref{slab}). It is evident that the predicted
net attenuations are very different in the two cases. We do not want
to discuss here the agreement of the result with respect to data, we
only wish to point out with one example the clearly different behavior
both in terms of trend and spread. Part of the difference is surely
due to the fact that with a slab the net attenuation is independent
from the bulge to disc ratio given by the SAM.  Instead, as discussed
by Tuffs et al. (2004), Pierini et al. (2004) and Driver et al.
(2007), the attenuation suffered by a dust-less bulge seen through the
disc can be very pronounced as a function of the viewing angle,
depending on the scale radii of stars and dust.  In {\gal} most bulges
host relevant star formation activity within dusty MCs, therefore in
these cases the total effective attenuation of the galaxy is dominated
by (spherically symmetric) bulges rather than by discs, and this
results in lower values and spread of the net attenuation as compared
to a slab distribution.
   
\section{Summary}
\label{final}

In this paper, we attempt to better understand and quantify the
results of detailed radiative transfer and dust models ({\gs}), with
the goal of assessing and improving the treatment of dust in
semi-analytic models of galaxy formation. We build several different
libraries of star formation/chemical enrichment histories, which we
couple with the {\gs} RT code. One library is based on the {\gal}
semi-analytic model, with galaxies extracted from two different
redshift intervals ($z<0.20$ and $2<z<3$). As a first choice, we keep
the composite geometry (bulge+disc) predicted by the SAM; we
additionally define a Control Library by considering only the dominant
galactic component (bulge or disc).  We then build a third library
using the chemical evolution code {\sc che\_evo} and requiring that
cold gas mass, metallicity and star formation rate are both
independent and uniformly distributed in the ($SFR$, $M_{\rm gas}$,
$Z$) space. For this Empirical library we consider a simpler geometry
(pure bulge or pure disc). We then interface the star formation
histories with the spectrophotometric code {\gs} to obtain the
corresponding synthetic extinguished and unextinguished SEDs.

We compute the intrinsic optical depth in the $V$-band $\tau_V^{GS}$
for each object, and we combine this estimate with different
prescriptions for the attenuation law and for the relative geometry of
stars and dust. We find that the {\gs} results are best reproduced by
a composite attenuation law as proposed by De Lucia \& Blaizot (2007),
which consists of a power-law in wavelength for the young stellar
population, as suggested by Charlot \& Fall (2000), and a Galactic
extinction curve plus a slab model for older stars.  However, we
notice that even a simpler model with a Galactic extinction law plus a
slab model is already a good approximation for the {\gs} results, if
the ``true'' value of $\tau_V^{GS}$ is known.

We then try to estimate the dependence of $\tau_V^{GS}$ on physical
properties of the model galaxies in the Empirical Library. We find
that $\tau_V^{GS}$ depends mostly on the gas metallicity, on the cold
gas mass and on the radius of the system. We provide fitting formulae
that are able to predict $\tau_V^{GS}$ over a large dynamical range
and with reasonable scatter. We then compare the predictions of the
fitting formulae with the intrinsic optical depth for SAM model
galaxies. We find that the proposed fitting formulae are able to
provide a reasonable fit to $\tau_V^{GS}$ in both redshift intervals.
The proposed fitting formulae are analogous to the Guiderdoni \&
Rocca-Volmerange (1987) and De Lucia \& Blaizot (2007) prescriptions;
however, we find different power law dependencies with respect to
those models.

We compare the {\gs} attenuations with prescriptions combining the De
Lucia \& Blaizot (2007) composite attenuation law with different
estimates of $\tau_V$. We conclude that our new fitting formula is
able to reproduce the distribution of attenuations and their scaling
with galaxy properties, such as stellar mass and specific star
formation rate, nearly as well as the model that uses the actual value
of $\tau_V^{GS}$ extracted from the RT libraries. The analytic recipes
from the literature (Guiderdoni \& Rocca-Volmerange 1987; De Lucia \&
Blaizot 2007) do reasonably well at low-z but give poor results at
high-z.

It is important to keep in mind that the SAM can provide many physical
properties of galaxies to {\gs}, such as star formation rates, gas
density and metallicity, and disc and bulge sizes, but it cannot
provide many important physical quantities of the ISM, like the
fraction of gas in molecular clouds, the timescale after which massive
stars emerge from the heavily obscuring molecular clouds or the
distribution of grain sizes. For the sake of simplicity, we have kept the
values of these parameters fixed for all model galaxies in our
libraries. Our choice reflects a parameter combination that has been
proven adequate for reproducing the properties of $z<3$ galaxies
(Silva et al., 1998; Fontanot et al., 2007). Little is known about the
redshift evolution of dust grain properties: a strong systematic
variation of these parameters may indeed affect the validity of our
fitting formulae. Observations of dust properties in the host galaxies
of high-z quasars (see e.g., Maiolino et al., 2004) suggest that the
intrinsic properties of dust grains (i.e. composition and dimension)
are indeed likely to change at $z>4$. This is probably due to the fact
that at such redshifts dust is mainly produced by Type II supernovae,
whereas at lower redshift the contribution from the envelopes of
evolved low-mass stars becomes dominant. Therefore we conclude that
our fitting formulae may be limited by this effect to $z<3$. Finally,
it is worth noting that even neglecting any variation of dust
properties with redshift, we still find a large dispersion in the
resulting attenuation laws: this result suggests that the effect of
the geometry and the complexity of star formation histories dominate
over the properties of the dust grains (see e.g. Granato et al.
2000).

Our results show that the use of a spectro-photometric code coupled
with a simple attenuation law may introduce a systematic shift in the
predicted magnitudes. This effect has to be taken into account as an
uncertainty associated with models when comparing their prediction to
observations. The best way to decrease this systematic error is to use
a full RT computation.

\section*{Acknowledgments}
The authors would like to thank Gabriella De Lucia for many enlighting
and detailed explanations of their model, and Patrick Jonsson, Richard
Tuffs and Cristina Popescu for stimulating discussions. Some of the
calculations were carried out on the PIA cluster of the
Max-Planck-Institut f\"ur Astronomie at the Rechenzentrum Garching.

{}


\begin{thebibliography}{}
\bibitem[]{ls7} Babbedge, T.~S.~R., et al.\ 2006, MNRAS, 370, 1159
\bibitem[{{Baugh}(2006)}]{baugh:06}
{Baugh} C.~M., 2006, Reports of Progress in Physics, 69, 3101
\bibitem[]{durha} Baugh C.M., Lacey C.G., Frenk C.S., Granato G.L.,
  Silva L., Bressan A., Benson A.J., Cole S., 2005, MNRAS, 356, 1191
\bibitem[]{eric2} Bell E., 2002, ApJ, 577, 150
\bibitem[]{ls38} Bell, E.~F., Baugh, C.~M., Cole, S., Frenk, C.~S., \&
  Lacey, C.~G.\ 2003, MNRAS, 343, 367
\bibitem[]{inc8} Bianchi, S., Ferrara, A., \& Giovanardi, C.\ 1996,
  ApJ, 465, 127
\bibitem[]{ls31} Blaizot, J., Guiderdoni, B., Devriendt, J.~E.~G.,
  Bouchet, F.~R., Hatton, S.~J., \& Stoehr, F.\ 2004, MNRAS, 352, 571
\bibitem[]{inc17} Boissier, S., Boselli, A., Buat, V., Donas, J., \&
  Milliard, B.\ 2004, A\&A, 424, 465
\bibitem[]{ls43} Bressan, A., Granato, G.~L., \& Silva, L.\ 1998,
  A\&A, 332, 135
\bibitem[]{gs2} Bressan A., Silva L., Granato G.L., 2002, A\&A, 392,
  377
\bibitem[]{bri04} Brinchmann, J. Charlot, S. White, S.D.M. Tremonti
  C., Kauffmann G., Heckman T., Brinkmann J., 2004, MNRAS, 351, 115
\bibitem[]{ls25} Bruzual A., G., Magris, G., \& Calvet, N.\ 1988, ApJ,
  333, 673
\bibitem[]{bc03} Bruzual, G., \& Charlot, S.\ 2003, MNRAS, 344, 1000
\bibitem[]{ls49} Buat, V., Marcillac, D., Burgarella, D., Le Floc'h,
  E., Takeuchi, T.~T., Iglesias-Par{\`a}mo, J., \& Xu, C.~K.\ 2007,
  A\&A, 469, 19
\bibitem[]{ls48} Buat, V., et al.\ 2005, ApJL, 619, L51
\bibitem[]{ls52} Burgarella, D., Le Floc'h, E., Takeuchi, T.~T.,
  Huang, J.~S., Buat, V., Rieke, G.~H., \& Tyler, K.~D.\ 2007, MNRAS,
  380, 986
\bibitem[]{ls53} Burgarella, D., Buat, V., \& Iglesias-P{\'a}ramo, J.\
  2005, MNRAS, 360, 1413
\bibitem[]{inc7} Byun, Y.~I., Freeman, K.~C., \& Kylafis, N.~D.\ 1994,
  ApJ, 432, 114
\bibitem[]{cal94} Calzetti D., Kinney A.L., Storchi-Bergmann T., 1994,
  ApJ, 492, 582
\bibitem[]{cal00} Calzetti D., Armus L., Bohlin R.C., Kinney A.L.,
  Koorneff J., Storchi-Bergmann T., 2000, ApJ, 533, 682
\bibitem[]{ls21} Calzetti, D.\ 2001, PASP, 113, 1449
\bibitem[]{ls18} Cardelli, J.~A., Clayton, G.~C., \& Mathis, J.~S.\
  1989, ApJ, 345, 245
\bibitem[]{cha03} Chapman S.C., Blain A.W., Ivison R.J., Smail I.R.,
  2003, Nat., 422, 695
\bibitem[]{ch5}Chapman, S.C., Blain, A.W., Smail, I.R., Ivison, R.J.,
  2005, ApJ, 622, 772
\bibitem[]{cef00} Charlot S., Fall S.M., 2000, ApJ, 539, 718
\bibitem[]{ls35} Cole, S., Lacey, C.~G., Baugh, C.~M., \& Frenk,
  C.~S.\ 2000, MNRAS, 319, 168
\bibitem[]{ls58} Cox, T.~J., Primack, J., Jonsson, P., \& Somerville,
  R.~S.\ 2004, ApJL, 607, L87
\bibitem[]{ls30} De Lucia, G., Kauffmann, G., \& White, S.~D.~M.\
  2004, MNRAS, 349, 1101
\bibitem[]{dlb} De Lucia G., \& Blaizot J.\ 2007, MNRAS, 375, 2
\bibitem[]{inc1b} de Vaucouleurs, G., de Vaucouleurs, A., Corwin,
  H.~G., Jr., Buta, R.~J., Paturel, G., \& Fouque, P.\ 1991, Volume
  1-3, XII, 2069 pp.~7 figs..~ Springer-Verlag Berlin Heidelberg New
  York,
\bibitem[]{ls13} Desert, F.-X., Boulanger, F., \& Puget, J.~L.\ 1990,
  A\&A, 237, 215
\bibitem[]{dgs99} Devriendt J.E.G., Guiderdoni B., Sadat, R., 1999,
  A\&A, 350, 381
\bibitem[]{ls36} Devriendt, J.~E.~G., \& Guiderdoni, B.\ 2000, A\&A,
  363, 851
\bibitem[]{dev98} Devriendt J.E.G., Sethi S.K., Guiderdoni B., Nath
  B.B., 1998, MNRAS, 298, 708
\bibitem[]{ls27} Dole, H., et al.\ 2001, A\&A, 372, 364
\bibitem[]{ls1} Dorschner, J., \& Henning, T.\ 1995, A\&ARv, 6, 271
\bibitem[]{drl03} Draine B.T., 2003, ApJ, 598, 1017
\bibitem[]{ls14} Draine, B.~T., \& Anderson, N.\ 1985, ApJ, 292, 494
\bibitem[]{inc14} Driver, S.~P., Popescu, C.~C., Tuffs, R.~J., Liske,
  J., Graham, A.~W., Allen, P.~D., \& de Propris, R.\ 2007, MNRAS,
  379, 1022
\bibitem[]{ls12} Dwek, E., et al.\ 1997, ApJ, 475, 565
\bibitem[]{dweva} Dwek E., Varosi F., 1996, In: {\it Unveiling the
    Cosmic Infrared Background}. AIP conference Proceedings, p.236
\bibitem[]{ls24} Efstathiou, A., \& Rowan-Robinson, M.\ 1995, MNRAS,
  273, 649
\bibitem[]{ls4} Elbaz, D., Cesarsky, C.~J., Chanial, P., Aussel, H.,
  Franceschini, A., Fadda, D., \& Chary, R.~R.\ 2002, A\&A, 384, 848
\bibitem[]{ls5} Elbaz, D., et al.\ 1999, A\&A, 351, L37
\bibitem[]{ferra} Ferrara A., Bianchi S., Cimatti A., Giovanardi C.,
  1999, ApJS, 123, 437
\bibitem[]{fioc} Fioc M. \& Rocca-Volmerange B., 1997, A\&A, 326, 950
\bibitem[]{ls42} Fontana, A., et al.\ 2006, A\&A, 459, 745
\bibitem[]{pap02} Fontanot F., Monaco P., Cristiani S., Tozzi P.,
  2006, MNRAS, 373, 1173
\bibitem[]{pap03} Fontanot F., Monaco P., Silva L., Grazian A., 2007,
  MNRAS, 382, 903
\bibitem[]{ls44} Fitzpatrick, E.\ 1989, IAUS, 135, 37
\bibitem[]{ls16} Fitzpatrick, E.~L.\ 1999, PASP, 111, 63
\bibitem[]{ls17} Fitzpatrick, E.~L., \& Massa, D.\ 2007, ApJ, 663, 320
\bibitem[]{inc2} Giovanelli, R., Haynes, M.~P., Salzer, J.~J., Wegner,
  G., da Costa, L.~N., \& Freudling, W.\ 1995, AJ, 110, 1059
\bibitem[]{ls50} Goldader, J.~D., Meurer, G., Heckman, T.~M., Seibert,
  M., Sanders, D.~B., Calzetti, D., \& Steidel, C.~C.\ 2002, ApJ, 568,
  651
\bibitem[]{ls23} Gordon, K.~D., Calzetti, D., \& Witt, A.~N.\ 1997,
  ApJ, 487
\bibitem[]{inc5} Graham, A.~W.\ 2001, MNRAS, 326, 543
\bibitem[]{gra00} Granato G., et al., 2000, ApJ, 542, 710
\bibitem[]{ls6} Gruppioni, C., Lari, C., Pozzi, F., Zamorani, G.,
  Franceschini, A., Oliver, S., Rowan-Robinson, M., \& Serjeant, S.\
  2002, MNRAS, 335, 831
\bibitem[]{grv87} Guiderdoni B. \& Rocca-Volmerange B., 1987, A\&A,
  186, 1
\bibitem[]{ls34} Guiderdoni, B., Hivon, E., Bouchet, F.~R., \& Maffei,
  B.\ 1998, MNRAS, 295, 877
\bibitem[]{jon06} Jonsson P., Cox T.J., Primack J.R., Somerville R.S.,
  2006, ApJ, 637, 255
\bibitem[]{ls47} Jonsson, P.\ 2006, MNRAS, 372, 2
\bibitem[]{lidr} Li A. \& Draine B.T, 2001, ApJ, 554, 778
\bibitem[]{maio} Maiolino, R., Schneider, R., Oliva, E., Bianchi, S.,
  Ferrara, A., Mannucci, F., Pedani, M., \& Roca Sogorb, M.\ 2004,
  Natur., 431, 533
\bibitem[]{inc3} Moriondo, G., Giovanelli, R., \& Haynes, M.~P.\ 1998,
  A \& A, 338, 795
\bibitem[]{ino05} Inoue, A.K., 2005, MNRAS, 359, 171
\bibitem[]{ino06} Inoue A.K., Buat V., Burgarella D., Panuzzo P.,
  Takeuchi T.T., Iglesias-Paramo J.I., 2006, MNRAS, 370, 380
\bibitem[]{c}Kauffmann, G., Colberg, J.M., Diaferio, A., \& White
  S.D.M., 1999, MNRAS, 303, 188
\bibitem[]{xi5} Kang X., Jing Y. P., Mo H. J., B\"orner G., 2005, ApJ,
  631 21
\bibitem[]{ls39} Kaviani, A., Haehnelt, M.~G., \& Kauffmann, G.\ 2003,
  MNRAS, 340, 739
\bibitem[]{ls32} Kitzbichler, M.~G., \& White, S.~D.~M.\ 2007, MNRAS,
  376, 2
\bibitem[]{kpng} Kong X., Charlot S., Brinchmann J., Fall S.M., 2004,
  349, 769
\bibitem[]{inc9} Kuchinski, L.~E., Terndrup, D.~M., Gordon, K.~D., \&
  Witt, A.~N.\ 1998, AJ, 115, 1438
\bibitem[]{inc15} Kylafis, N.~D., \& Bahcall, J.~N.\ 1987, ApJ, 317,
  637
\bibitem[]{ls33} Lacey, C., Guiderdoni, B., Rocca-Volmerange, B., \&
  Silk, J.\ 1993, ApJ, 402, 1 5
\bibitem[]{ls37} Lacey, C.~G., Baugh, C.~M., Frenk, C.~S., Silva, L.,
  Granato, G.~L., \& Bressan, A.\ 2007, ArXiv e-prints, 704,
  arXiv:0704.1562
\bibitem[]{ls8} Le Floc'h, E., et al.\ 2005, ApJ, 632, 169
\bibitem[]{stard} Leitherer C. et al., 1999, ApJS, 123, 3
\bibitem[]{ls57} Lucy, L.~B., Danziger, I.~J., Gouiffes, C., \&
  Bouchet, P.\ 1991, in Supernovae ed. S.E.~Woosley, (Springer-Verlag,
  New York), 82
\bibitem[]{at2} Hatton S., Devriendt J.E.G., Ninin S., Bouchet F.R.,
  Guiderdoni, B., \& Vibert, D.\ 2003, MNRAS, 343, 75
\bibitem[]{ls9} Hauser, M.~G., et al.\ 1998, ApJ, 508, 25
\bibitem[]{ls20} Hauser, M. G., \& Dwek, E. 2001, ARA\&A, 39, 249
\bibitem[]{at} Hughes, D.~H., et al.\ 1998, Nat, 394, 241
\bibitem[]{ls54} Iglesias-P{\'a}ramo, J., et al.\ 2007, ApJ, 670, 279
\bibitem[]{rsfrS} Madgwick D.S., Somerville R., Lahav O., Ellis R.,
  2003, MNRAS, 343, 871
\bibitem[]{inc6} Masters, K.~L., Giovanelli, R., \& Haynes, M.~P.\
  2003, AJ, 126, 158
\bibitem[]{ls29} Mathis, H., Lemson, G., Springel, V., Kauffmann, G.,
  White, S.~D.~M., Eldar, A., \& Dekel, A.\ 2002, MNRAS, 333, 739
\bibitem[]{ls15} Mathis, J.~S.\ 1990, ARA\&A, 28, 37
\bibitem[]{mat83} Mathis J.S., Mezger P.G., Panagia N., A\&A, 1983,
  128, 212
\bibitem[]{ls51} Meurer, G.~R., Heckman, T.~M., \& Calzetti, D.\ 1999,
  ApJ, 521, 64
\bibitem[]{mmw} Mo, H.~J., Mao, S., \& White, S.~D.~M.\ 1998, MNRAS,
  295, 319
\bibitem[]{inc13} M{\"o}llenhoff, C., Popescu, C.~C., \& Tuffs, R.~J.\
  2006, A\&A, 456, 941
\bibitem[]{bs} Monaco P., 2004, MNRAS, 352, 181
\bibitem[]{a} Monaco P., \& Fontanot F.\ 2005, MNRAS, 359, 283
\bibitem[]{icl}Monaco P., Murante G., Borgani S., Fontanot F.,
  2006, ApJL, 652, 89
\bibitem[]{pap01} Monaco P., Fontanot F. \& Taffoni G., 2007, MNRAS,
  375, 1189
\bibitem[]{ls28} Nagashima, M., Totani, T., Gouda, N., \& Yoshii, Y.\
  2001, ApJ, 557, 505
\bibitem[]{ls45} Natta, A., \& Panagia, N.\ 1984, ApJ, 287, 228
\bibitem[]{pan} Panuzzo P., Bressan A., Granato G.~L., Silva L.,
  Danese L., 2003, A\&A, 409, 99
\bibitem[]{panuz} Panuzzo P., Granato G.L., Buat V., Inoue A.K., Silva L.,
  Iglesias-Paramo J., Bressan A., 2007, MNRAS, 1444
\bibitem[]{inc10} Pierini, D., Gordon, K.~D., Witt, A.~N., \& Madsen,
  G.~J.\ 2004, ApJ, 617, 1022
\bibitem[]{pop00} Popescu C.C., Misiriotis A., Kylafis N.D., Tuffs
  R.J., Fischera J., 2000, A\&A, 362, 138
\bibitem[]{poptu} Popescu C.C. \& Tuffs R.J., 2002, MNRAS, 335, L41
\bibitem[]{inc12b} Popescu, C.~C., et al.\ 2005, ApJL, 619, L75
\bibitem[]{ls10} Puget, J.-L., Abergel, A., Bernard, J.-P., Boulanger,
  F., Burton, W.~B., Desert, F.-X., \& Hartmann, D.\ 1996, A\&A, 308,
  L5
\bibitem[]{inc11} Rocha, M., Jonsson, P., Primack, J.~R., \& Cox,
  T.~J.\ 2008, MNRAS, 383, 1281
\bibitem[]{ls19} Rowan-Robinson, M.\ 1986, MNRAS, 219, 737
\bibitem[]{ls3} Sanders, D.~B., \& Mirabel, I.~F.\ 1996, ARA\&A, 34,
  749
\bibitem[]{salpt} Salpeter E.E., 1955, ApJ, 121, 61
\bibitem[]{ls40} Silva, L., De Zotti, G., Granato, G.~L., Maiolino,
  R., \& Danese, L.\ 2005, MNRAS, 357, 1295
\bibitem[]{s2} Silva L., 1999, PhD thesis, SISSA, Trieste
\bibitem[]{sil98} Silva L., Granato G.L., Bressan A., Danese L.,
  1998, ApJ, 509, 103
\bibitem[]{sc9}Smail, I., Ivison, R., Blain, A., 1997, ApJ, 490, L5
\bibitem[]{sc10} Smail I., Ivison R., Blain A., Kneib J.-P., 2002,
  MNRAS, 331, 495
\bibitem[]{ls2} Soifer, B.~T., \& Neugebauer, G.\ 1991, AJ, 101, 354
\bibitem[]{rsjp99}Somerville R.~S. \& Primack J.~R., 1999, MNRAS, 310,
  1087
\bibitem[]{rache} Somerville R.~S. et al. 2007
\bibitem[]{tuffs} Tuffs R.J., Popescu C.C., V\"olk H.J., Kylafis N.D.,
  Dopita M.A., 2004, A\&A, 419, 821
\bibitem[]{inc4} Tully, R.~B., Pierce, M.~J., Huang, J.-S., Saunders,
  W., Verheijen, M.~A.~W., \& Witchalls, P.~L.\ 1998, AJ, 115, 2264
\bibitem[]{ls46} V{\'a}rosi, F., \& Dwek, E.\ 1999, ApJ, 523, 265
\bibitem[]{veg} Vega O., Silva L., Panuzzo P., Bressan A., Granato
  G.~L., Chavez M., 2005, MNRAS, 364, 1286
\bibitem[]{ls41} Viola, M., Monaco, P., Borgani, S., Murante, G., \&
  Tornatore, L.\ 2007, MNRAS, 1114
\bibitem[]{ls55} Xu, C.~K., et al.\ 2007, ArXiv Astrophysics e-prints,
  arXiv:astro-ph/0701737
\bibitem[]{ls22} Wang, B., \& Heckman, T.~M.\ 1996, ApJ, 457, 645
\bibitem[]{wf} White, S.~D.~M., \& Frenk, C.~S.\ 1991, ApJ, 379, 52
\bibitem[]{ls26} Witt, A.~N., Thronson, H.~A., Jr., \& Capuano, J.~M.,
  Jr.\ 1992, ApJ, 393, 611
\bibitem[]{inc16} Xilouris, E.~M., Byun, Y.~I., Kylafis, N.~D.,
  Paleologou, E.~V., \& Papamastorakis, J.\ 1999, A\&A, 344, 868
\bibitem[]{ls11} Zubko, V., Dwek, E., \& Arendt, R.~G.\ 2004, ApJS,
  152, 211
\end{thebibliography}
\end{document}